\documentstyle[12pt,psfig]{article}
\makeatletter
\def\@cite#1#2{{[{#1}]\if@tempswa\typeout
{IJCGA warning: optional citation argument
ignored: `#2'} \fi}}

\newcount\@tempcntc
\def\@citex[#1]#2{\if@filesw\immediate\write\@auxout{\string\citation{#2}}\fi
  \@tempcnta\z@\@tempcntb\m@ne\def\@citea{}\@cite{\@for\@citeb:=#2\do
    {\@ifundefined
       {b@\@citeb}{\@citeo\@tempcntb\m@ne\@citea\def\@citea{,}{\bf ?}\@warning
       {Citation `\@citeb' on page \thepage \space undefined}}%
    {\setbox\z@\hbox{\global\@tempcntc0\csname b@\@citeb\endcsname\relax}%
     \ifnum\@tempcntc=\z@ \@citeo\@tempcntb\m@ne
       \@citea\def\@citea{,}\hbox{\csname b@\@citeb\endcsname}%
     \else
      \advance\@tempcntb\@ne
      \ifnum\@tempcntb=\@tempcntc
      \else\advance\@tempcntb\m@ne\@citeo
      \@tempcnta\@tempcntc\@tempcntb\@tempcntc\fi\fi}}\@citeo}{#1}}
\def\@citeo{\ifnum\@tempcnta>\@tempcntb\else\@citea\def\@citea{,}%
  \ifnum\@tempcnta=\@tempcntb\the\@tempcnta\else
   {\advance\@tempcnta\@ne\ifnum\@tempcnta=\@tempcntb \else
\def\@citea{--}\fi
    \advance\@tempcnta\m@ne\the\@tempcnta\@citea\the\@tempcntb}\fi\fi}
\makeatother
\newenvironment{Eqnarray}%
     {\arraycolsep 0.14em\begin{eqnarray}}{\end{eqnarray}}

\def\simlt{\stackrel{<}{{}_\sim}}
\def\simgt{\stackrel{>}{{}_\sim}}
\def\be{\begin{equation}}
\def\ee{\end{equation}}
\def\bear{\be\begin{array}}
\def\eear{\end{array}\ee}
\def\bea{\begin{Eqnarray}}
\def\eea{\end{Eqnarray}}

\def\lsim{\mathrel{\raise.3ex\hbox{$<$\kern-.75em\lower1ex\hbox{$\sim$}}}}
\def\gsim{\mathrel{\raise.3ex\hbox{$>$\kern-.75em\lower1ex\hbox{$\sim$}}}}
\def\ifmath#1{\relax\ifmmode #1\else $#1$\fi}
\def\ls#1{\ifmath{_{\lower1.5pt\hbox{$\scriptstyle #1$}}}}

\def\beq{\begin{equation}}
\def\eeq{\end{equation}}
\def\beqa{\begin{Eqnarray}}
\def\eeqa{\end{Eqnarray}}

\def\simlt{\stackrel{<}{{}_\sim}}  \def\simgt{\stackrel{>}{{}_\sim}}

\def\baselinestretch{1}
\begin{document}
\def\IJMPA #1 #2 #3 {{\sl Int.~J.~Mod.~Phys.}~{\bf A#1}\ (19#2) #3$\,$}
\def\MPLA #1 #2 #3 {{\sl Mod.~Phys.~Lett.}~{\bf A#1}\ (19#2) #3$\,$}
\def\NPB #1 #2 #3 {{\sl Nucl.~Phys.}~{\bf B#1}\ (19#2) #3$\,$}
\def\PLB #1 #2 #3 {{\sl Phys.~Lett.}~{\bf B#1}\ (19#2) #3$\,$}
\def\PR #1 #2 #3 {{\sl Phys.~Rep.}~{\bf#1}\ (19#2) #3$\,$}
\def\JHEP #1 #2 #3 {{\sl JHEP}~{\bf #1}~(19#2)~#3$\,$}
\def\PRD #1 #2 #3 {{\sl Phys.~Rev.}~{\bf D#1}\ (19#2) #3$\,$}
\def\PTP #1 #2 #3 {{\sl Prog.~Theor.~Phys.}~{\bf #1}\ (19#2) #3$\,$}
\def\PRL #1 #2 #3 {{\sl Phys.~Rev.~Lett.}~{\bf#1}\ (19#2) #3$\,$}
\def\RMP #1 #2 #3 {{\sl Rev.~Mod.~Phys.}~{\bf#1}\ (19#2) #3$\,$}
\def\ZPC #1 #2 #3 {{\sl Z.~Phys.}~{\bf C#1}\ (19#2) #3$\,$}
\def\PPNP#1 #2 #3 {{\sl Prog. Part. Nucl. Phys. }{\bf #1} (#2) #3$\,$}

\catcode`@=11
\newtoks\@stequation
\def\subequations{\refstepcounter{equation}%
\edef\@savedequation{\the\c@equation}%
  \@stequation=\expandafter{\theequation}
  \edef\@savedtheequation{\the\@stequation}
  \edef\oldtheequation{\theequation}%
  \setcounter{equation}{0}%
  \def\theequation{\oldtheequation\alph{equation}}}
\def\endsubequations{\setcounter{equation}{\@savedequation}%
  \@stequation=\expandafter{\@savedtheequation}%
  \edef\theequation{\the\@stequation}\global\@ignoretrue

\noindent}
\catcode`@=12
\begin{titlepage}

\title{{\bf Large mixing angles for neutrinos \\
from infrared fixed points }}
\vskip2in
\author{ 
{\bf J.A. Casas$^{1,2}$\footnote{\baselineskip=16pt E-mail: {\tt
alberto@makoki.iem.csic.es}}}, 
{\bf J.R. Espinosa$^{2,3}$\footnote{\baselineskip=16pt E-mail: {\tt
espinosa@makoki.iem.csic.es}}} and 
{\bf I. Navarro$^{4}$\footnote{\baselineskip=16pt E-mail: {\tt
ignacio.navarro@durham.ac.uk}}}\\ 
\hspace{3cm}\\
 $^{1}$~{\small IEM (CSIC), Serrano 123, 28006 Madrid, Spain}
\hspace{0.3cm}\\
$^{2}$~{\small IFT C-XVI, UAM., Cantoblanco, 28049 Madrid, Spain }
\hspace{0.3cm}\\
$^{3}$~{\small IMAFF (CSIC), Serrano 113 bis, 28006 Madrid, Spain}
\hspace{0.3cm}\\
 $^{4}$~{\small IPPP,  University of Durham, DH1 3LE, Durham, UK}} 
\date{} 
\maketitle 
\def\baselinestretch{1.15} 
\begin{abstract}
\noindent
Radiative amplification of neutrino mixing angles may explain the large
values required by solar and atmospheric neutrino oscillations.
Implementation of such mechanism in the Standard Model and many of its
extensions (including the Minimal Supersymmetric Standard Model) to
amplify the solar angle, the atmospheric or both requires (at least two)
quasi-degenerate neutrino masses, but is not always possible. When it is,
it involves a fine-tuning between initial conditions and radiative
corrections. In supersymmetric models with neutrino masses
generated through the K\"ahler potential, neutrino mixing angles can
easily be driven to large values at low energy as they approach infrared
pseudo-fixed points at large mixing (in stark contrast with conventional
scenarios, that have infrared pseudo-fixed points at zero mixing). In
addition, quasi-degeneracy of neutrino masses is not always required.
 \end{abstract}

\thispagestyle{empty} \leftline{June 2003} \leftline{}

\vskip-20cm \rightline{} 
\rightline{IEM-FT/231-03}
\rightline{IFT-UAM/CSIC-03-20} 
\rightline{IPPP/03/38}
\rightline{DCPT/03/76}
\rightline{hep-ph/0306243} \vskip3in

\end{titlepage}
\setcounter{footnote}{0} \setcounter{page}{1}
\newpage
\baselineskip=20pt

\section{Introduction}

The experimental study of flavour non-conservation in diverse types of 
neutrino fluxes (solar, atmospheric and "man-made") has produced in recent 
years considerable evidence in favour of oscillations among massive 
neutrinos \cite{PMNS}. Theoretically, the most economic scenario to 
accomodate the 
data (or at least the more firmly stablished data, therefore leaving aside 
the LSND anomaly \cite{LSND}) assumes that the left-handed neutrinos of 
the Standard Model (SM) acquire Majorana masses through a dimension-5 
operator \cite{Weinberg}, which is the 
low-energy trace of lepton-number violating physics at much higher energy 
scales (the simplest example being the see-saw \cite{seesaw}).

Neutrino masses are then described by a $3\times 3$ mass 
matrix ${\cal M}_\nu$ that is diagonalized
by the  PMNS \cite{PMNS} unitary matrix $V$:
\be \label{Udiag}
V^T{\cal M}_\nu V={\mathrm diag}(m_1,m_2,m_3)\ .
\ee
The masses $m_i$ are real (not necessarily positive) numbers.  Following a
standard convention we denote by $m_3$ the most split eigenvalue
and choose $|m_1|\leq |m_2|$. For later use we define
the quantities
\be
\label{deltadef0}
\Delta m^2_{ij}\equiv m_i^2-m_j^2\ ,\;\;\;
\Delta_{ij}\equiv\frac{m_i-m_j}{m_i+m_j}\ ,\;\;\;
\nabla_{ij}\equiv\frac{m_i+m_j}{m_i-m_j}\ .
\ee
The latter plays an important r\^ole in the RG evolution of $V$. 
For simplicity we set CP-violating phases to zero throughout the paper, 
so $V$ can be parametrized by three 
succesive rotations  as
\be 
\label{Vdef}
V=R_{23}(\theta_1)R_{31}(\theta_2)R_{12}(\theta_3)= 
\pmatrix{c_2c_3 & c_2s_3 & s_2\cr
-c_1s_3-s_1s_2c_3 & c_1c_3-s_1s_2s_3 & s_1c_2\cr
s_1s_3-c_1s_2c_3 & -s_1c_3-c_1s_2s_3 & c_1c_2\cr}\ ,
\ee
where $s_i\equiv \sin \theta_i$, $c_i\equiv \cos \theta_i$.  

The experimental information on the neutrino
sector is the following. For
the CHOOZ angle: $\sin^2\theta_2< 0.052$; for the atmospheric neutrino 
parameters: $1.5\times 10^{-3} <  \Delta m^2_{atm}/{\rm eV}^2 < 3.9\times 
10^{-3}$ and $0.45 < \tan^2\theta_1<2.3$;
and  for the solar ones (with the MSW
mechanism \cite{MSW} at work): $5.4\times 10^{-5} <  \Delta 
m^2_{sol}/{\rm eV}^2 < 10^{-4}$ or 
$1.4\times 10^{-4} <  \Delta m^2_{sol}/{\rm eV}^2 < 1.9\times 10^{-4}$
and $0.29 < \tan^2\theta_3<0.82$. 
These (3$\sigma$ CL) ranges arise from the global statistical analysis 
\cite{status}
of many experimental data coming from neutrino fluxes
of accelerator (K2K \cite{K2K}), reactor (CHOOZ, KAMLAND,... 
\cite{CHOOZ,KamLAND}),
atmospheric (SK, MACRO, SOUDAN-2 \cite{SK}--\cite{SOUDAN2}) and solar 
(Kamiokande, SK, 
SNO,...\cite{Kamiokande}--\cite{GNO}) origin.
The smallness of $\theta_2$ and the hierarchy of mass splittings
implies that the oscillations of atmospheric and
solar neutrinos are dominantly two-flavour oscillations, described by a
single mixing angle and mass splitting: $\theta_{atm}\equiv\theta_1$,
$\Delta m^2_{atm}\equiv\Delta m^2_{31}\sim \Delta m^2_{32}$ and  
$\theta_{sol}\equiv\theta_3$, $\Delta m^2_{sol}\equiv\Delta m^2_{21}$.

Concerning the overall scale of neutrino masses,
the non-observation of neutrinoless double $\beta$-decay requires 
the $ee$ element of ${\cal M}_\nu$ to satisfy~\cite{HeidelbergMoscow}
\be
{\cal M}_{ee} \equiv
\big\vert  m_{1}\,c_2^2c_3^2 +
m_{2}\, c_2^2s_3^2 + m_{3}\, s_2^2
\big\vert\simlt 0.27\ {\rm eV}\ .
\label{B}
\ee
In addition,
Tritium $\beta$-decay experiments \cite{triti}, set the
bound  $m_i < 2.2$ eV for any mass eigenstate with a significant $\nu_e$ 
component. Finally, astrophysical observations of great cosmological 
importance, like those of 2dFGRS \cite{2dFGRS} and especially WMAP 
\cite{WMAP}  
set the limit $\sum_i |m_i|<0.69$ eV. This still allows three  
possibilities for the neutrino spectrum: hierarchical 
($m_1^2\ll m_2^2\ll m_3^2$), inverted-hierarchical 
($m_1^2\simeq m_2^2\gg m_3^2$) and quasi-degenerate
($m_1^2\simeq m_2^2\simeq m_3^2$).

The nearly bi-maximal structure of the neutrino mixing matrix, $V$, is
very different from that of the quark sector, where all the mixings are
small. An attractive possibility to explain this is that some neutrino
mixings are radiatively enhanced, i.e. are initially small and get large
in the Renormalization Group (RG) running from high to low energy (RG
effects on neutrino parameters have been discussed in
\cite{RG1}--\cite{mohapatra}). This amplification effect has been 
considered at
large in the literature 
\cite{RG2,CEIN4}, \cite{Tanimoto}--\cite{mohapatra}, but 
quite often
the analyses were incomplete or even incorrect.

In this paper we carefully examine this mechanism for radiative
amplification of mixing angles, paying particular attention to {\it 1)} a
complete treatment of all neutrino parameters (to ensure that not only
mixing angles but also mass splittings agree with experiment at low
energy) and {\it 2)} the fine-tuning price of amplification. We perform
this analysis in conventional scenarios, like the Standard Model (SM) or
the Minimal Supersymmetric Standard Model (MSSM) and confront them with
unconventional supersymmetric scenarios, proposed recently, in which
neutrino masses originate in the K\"ahler potential 
\cite{PRL}.\footnote{We restrict 
our analysis to the simplest low-energy effective models for neutrino 
masses, with no other assumptions on the physics at high-energy. We 
therefore do not discuss RG effects in see-saw scenarios, 
which have been 
considered previously, {\it e.g.} in \cite{CEIN1,CEIN4,Antusch}.} 
The sources of
neutrino masses in both types of scenarios and their renormalization group
equations (RGEs) are reviewed in Section~2, which also includes a generic
discussion of the presence of infrared pseudo-fixed points (IRFP) in the
running of the mixing angles. 
Section~3 is devoted to the radiative 
amplification of mixing angles in the conventional scenarios (SM and 
MSSM): we start with an illustrative toy model of only two flavours 
and later we apply the mechanism first to the amplification of the solar 
angle, then to the atmospheric angle and finally to the 
simultaneous amplification of both.
Section~4 deals with the amplification of the mixing angles in the 
unconventional supersymmetric model which looks quite promising due to 
its peculiar RG features. We collect some conclusions in 
Section~5. Appendix~A contains quite generic renormalization group 
equations for neutrino masses and mixing angles, while Appendix~B
presents  renormalization group
equations for generic non-renormalizable operators in the K\"ahler 
potential (like the ones responsible for neutrino masses in the 
unconventional scenario discussed in this paper).

\section{Sources of neutrino masses and RGEs}

\subsection{Conventional SM and MSSM}

In the SM the lowest order operator that generates Majorana
neutrino masses is
\be   
\delta{\cal L}=-\frac{1}{4M}\lambda_{\alpha\beta} (H\cdot L_\alpha)(H\cdot 
L_\beta)
+ {\mathrm h.c.},
\label{D5OSM}
\ee
where $H$ is the SM Higgs doublet, $L_\alpha$ is the lepton doublet of the 
$\alpha^{th}$ family, $\lambda_{\alpha\beta}$ is a (symmetric) matrix in 
flavor space 
and $M$ is the scale of the new physics  that violates lepton
number, L. After electroweak symmetry breaking, the neutrino mass matrix 
is ${\cal M}_\nu=\lambda v^2/(4M)$,
where $v=246$ GeV (with this definition $\lambda$ and ${\cal M}_{\nu}$
obey the same RGE).  This scheme can be easily made supersymmetric.
The standard SUSY framework has an operator
\be  
\delta{W}=-\frac{1}{4M}\lambda_{\alpha\beta} (H_2\cdot 
L_\alpha)(H_2\cdot L_\beta),
\label{D5OMSSM} 
\ee
in the superpotential $W$, giving ${\cal M}_\nu= \lambda \langle H_2^0
\rangle^2/(2M) = \lambda v^2\sin^2 \beta/(4M)$ (with $\tan \beta=\langle 
H^0_2
\rangle/\langle H^0_1 \rangle$).  Both in the SM and the MSSM the 
energy-scale evolution of ${\cal M}_\nu$ is governed by a RGE 
\cite{RG1}--\cite{RG4}
of the form ($t=\log Q$):
\be  \frac{d{\cal M}_\nu}{dt}=-(u_M{\cal M}_\nu 
+ c_M P_E {\cal M}_\nu + c_M {\cal M}_\nu P_E^T)\ ,
\label{RGM} 
\ee
where $P_E\equiv Y_E Y_E^\dagger/(16\pi^2)$ with $Y_E$  the matrix of 
leptonic Yukawa couplings. The
model-dependent quantities $u_M$ and $c_M$ are given in Appendix~A.
Notice that the
non-renormalizable operator of Eq.~(\ref{D5OSM}) [Eq.~(\ref{D5OMSSM})
for the SUSY case]  is the only L-violating operator  in the
effective theory, thus its presence in the right-hand side of
Eq.~(\ref{RGM}). The term $u_M{\cal M}_\nu$ gives a family-universal scaling 
of
${\cal M}_\nu$ which does not affect its texture, while the interesting 
non family-universal corrections, that can affect the neutrino mixing 
angles, appear through the matrix $P_E$. 

A very important difference between the SM and the MSSM is the value of 
the squared tau-Yukawa coupling in $P_E$. One has:
\be
y_\tau^2=\left\{
\begin{array}{ll}
{\displaystyle{{2m_\tau^2\over v^2}}}\ ,& ({\mathrm SM})\vspace{5mm}\\
{\displaystyle{{2m_\tau^2\over v^2\cos^2\beta}}}=
{\displaystyle{{2m_\tau^2\over v^2}}}(1+\tan^2\beta)
\ .& ({\mathrm MSSM})
\end{array}
\right.
\ee
Therefore, RG effects can be much larger in the MSSM for sizeable 
$\tan\beta$.

\subsection{Neutrino masses from the K\"ahler potential}

Operators that 
violate L-number in the K\"ahler potential, $K$, 
offer an alternative supersymmetric source of neutrino masses \cite{PRL}. 
The lowest-dimension (non-renormalizable) operators of this kind (that 
respect $R$-parity) are
\bea
\label{KL}
\delta K = {1\over 2M^2}\kappa_{\alpha\beta}(L_\alpha\cdot 
H_2)(L_\beta\cdot
\overline{H}_1) +{1\over 4M^2}\kappa'_{\alpha\beta}(L_\alpha\cdot 
\overline{H}_1)
(L_\beta\cdot \overline{H}_1)+ {\mathrm h.c.}\ ,  
\eea
where 
$\kappa$, 
$\kappa'$ are dimensionless matrices in flavour space
and $\overline{H}_1=-i\sigma_2 H_1^*$. While $\kappa'$ is a symmetric 
matrix, $\kappa$ may
contain a symmetric and an antisymmetric part: $\kappa\equiv
\kappa_S+\kappa_A$.  These operators give a neutrino mass matrix 
\cite{PRL}
\bea
\label{kahlernu}
{\cal M}_\nu=\frac{\mu v^2}{M^2} \left[ \kappa_S\sin^2\beta
+\kappa'\sin\beta\cos\beta\right]\ ,
\eea
where $\mu$ is the SUSY Higgs mass in the superpotential, $W\supset \mu
H_1 \cdot H_2$. If $W$ also contains the conventional operator 
(\ref{D5OMSSM}),
the contribution (\ref{kahlernu}) is negligible in comparison (by a
factor $\mu/M\ll 1$). As shown
in \cite{PRL} there are symmetries that can forbid the operator 
(\ref{D5OMSSM})
and leave (\ref{kahlernu}) as the only source of neutrino masses. This is
our assumption for this scenario. Moreover, as we discuss below,
interesting new effects appear through the matrix $\kappa$. Therefore we
focus on it as the main source of neutrino masses and set $\kappa'=0$.
This can also be the result of some symmetry \cite{PRL} or be a good
approximation if the mass that suppresses the
$\kappa'$ operator is much larger than that for $\kappa$. Another
possibility is that $\tan\beta$ is large (as is common in this context),
in which case the contribution of $\kappa'$ to neutrino masses is
suppressed by $\sim 1/\tan\beta$.
Finally, note that, due to the extra suppression factor $\mu/M$, in 
this 
scenario $M$ is much smaller than in the conventional case.

Appendix~B presents the RGEs for some non-renormalizable couplings in the 
K\"ahler potential, of which $\kappa$ and $\kappa'$ are particular 
examples. The  matrix
$\kappa'$ obeys a RGE of the form (\ref{RGM}) and therefore behaves like
the conventional case, while the RGE for
$\kappa$  has a remarkable structure \cite{PRL}
\bea
\label{kapp}
{d\kappa\over dt}&=&u\kappa+ P_E\kappa-\kappa
P_E^T+2(P_E\kappa-\kappa^T P_E^T)\ ,  \eea
where $16\pi^2 u={\mathrm Tr}(3Y_U^\dagger Y_U+ 3Y_D^\dagger Y_D+Y_E^\dagger
Y_E)-3g_2^2-g_1^2$. Here $Y_{U(D)}$ is the matrix of up (down) quark 
Yukawa couplings while $g_2$ and $g_1$ are the $SU(2)_L$ and $U(1)_Y$ 
gauge couplings, respectively.
\begin{figure}[t]
\centerline{
\psfig{figure=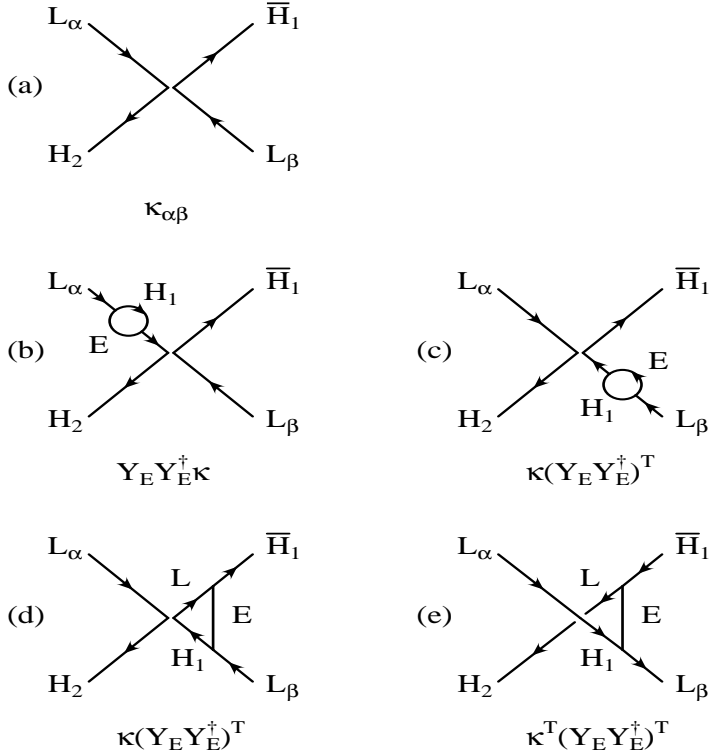,height=10cm,width=5cm,bbllx=7.cm,%
bblly=6.cm,bburx=14.cm,bbury=24.cm}}
\caption{
\noindent{\footnotesize The coupling
$\kappa_{\alpha\beta}$ and one-loop texture-changing radiative
corrections to it. The lines labelled E represent right-handed 
(charged) lepton superfields.}}
\label{fig:diag}
\end{figure}

Besides the usual universal piece, $u\kappa$, there are two different
terms that can change the texture of $\kappa$ and are therefore the
most interesting. The first, $P_E\kappa-\kappa
P_E^T$, decomposes in a symmetric and an antisymmetric part. In that
order:
\be  P_E\kappa-\kappa P_E^T=(P_E\kappa_A-\kappa_A P_E^T)
+(P_E\kappa_S-\kappa_S P_E^T)\ .  \ee
The second texture-changing term, $2(P_E\kappa-\kappa^T P_E^T)$, is
antisymmetric and, therefore, contributes only to the RG evolution of
$\kappa_A$, the antisymmetric part of $\kappa$.

The 
diagrammatic 
origin of these contributions is explained with the
help  of figure~\ref{fig:diag}. Diagram~(a) is a tree-level supergraph 
for the coupling $\kappa_{\alpha\beta}$. The order of subindices 
is 
important: 
$L_\alpha$ is $SU(2)$-contracted with $H_2$; $L_\beta$ with
$\overline{H}_1$. This is depicted in figure~\ref{fig:diag} by the two
"branches" of the vertex, with arrows indicating the order in the
$SU(2)$  product. We do not show the one-loop supergraphs that contribute 
to
the universal  renormalization of $\kappa_{\alpha\beta}$ but focus on
those that can  change its texture. 
Diagrams (b) and (c)
 renormalize  $\kappa_{\alpha\beta}$ through the anomalous
dimensions of the leptonic  legs, $L_{\alpha}$, $L_{\beta}$. These
kinds of diagrams are proportional to $P_E\kappa$ and $\kappa P_E^T$, as
indicated, and are also present when  neutrino mass operators arise
from the superpotential. They contribute a  $P_E\kappa+\kappa P_E^T$
piece to the  renormalization of $\kappa$. 
Diagrams (d) and (e) are non zero
only because $\kappa$ involves chiral and anti-chiral fields.  Similar
vertex corrections are absent for the neutrino mass operator
in $W$, which involves only
chiral fields, and is protected by SUSY non-renormalization
theorems. Diagram (d) gives a contribution similar to that coming from
diagram (c) but twice as large and with opposite sign. The net effect
is  to change $P_E\kappa+\kappa P_E^T$ [from (b)+(c)] into
$P_E\kappa-\kappa  P_E^T$. This is the origin of the first term in the
RGE (\ref{kapp}).  Finally, diagram (e) gives only a correction to the
operator  $(L_\alpha\cdot L_\beta)(\overline{H}_1\cdot H_2)$, which is
the  antisymmetric part of $\kappa_{\alpha\beta}$ by virtue of the
identity $r_{\alpha\beta}(L_\alpha\cdot
L_\beta)(\overline{H}_1\cdot H_2)=
(r_{\beta\alpha}-r_{\alpha\beta})(L_\alpha\cdot H_2)
(L_\beta\cdot\overline{H}_1)$ and this is responsible for the
last term $2(P_E\kappa-\kappa^T P_E^T)$  in (\ref{kapp}).

In order to show more clearly the structure of the RGE for  $\kappa$,
Eq.~(\ref{kapp}), it is  convenient to split it in two: one for the
symmetric part, $\kappa_S$, that  is directly responsible for neutrino
masses, and another for the  antisymmetric part, $\kappa_A$, that does
not contribute to neutrino masses. One gets
\bea
\label{kappaS}
{d\kappa_S\over dt}&=&u\kappa_S +P_E\kappa_A-\kappa_AP_E^T\ ,\\
{d\kappa_A\over dt}&=&u\kappa_A +P_E\kappa_S- \kappa_S P_E^T
+2(P_E\kappa- \kappa^T P_E^T)\ .
\label{kappaA}
\eea
As explained in \cite{PRL}, the RGE for $\kappa_S$ has the remarkable
property of being dependent of $\kappa_S$ itself only through the
universal piece. We have shown in more detail here how this arises
from a  cancellation involving corrections that are only present in
supersymmetry  for couplings in the K\"ahler
potential. Non-supersymmetric  two-Higgs-doublet models also have
vertex corrections, but there is no  such cancellation there. 
Some interesting  implications that follow from
the RGEs (\ref{kappaS}, \ref{kappaA}) were  presented in 
\cite{PRL}.\footnote{For 
instance, if initially $\kappa_S=0$ the whole neutrino mass matrix is 
generated as a radiative effect through (\ref{kappaS}). Such matrix has 
precisely the texture of the Zee model \cite{zee} (actually, this 
possibility can be understood as the supersymmetrization of the Zee 
model).} In this paper we will study in detail the   
possibilities they offer for amplifying neutrino mixing angles in a 
natural  way.

\subsection{Infrared pseudo-fixed points (IRFP) for mixing angles}

Equations (\ref{RGM}) and (\ref{kappaS}) detail how ${\cal M}_\nu$
receives a non-universal RG perturbation which is in general modest ($P_E$
is dominated by $y_\tau^2$, which is very small, unless $\tan \beta\simgt
50$).  However, when ${\cal M}_\nu$ has (quasi-)degenerate eigenvalues
($m_i\simeq m_j$), even small perturbations can cause large effects in the
eigenvectors ({\it i.e.} in the form of $V$). This can be easily
understood: for exact degeneracy there is an ambiguity in the choice of
the associated eigenvectors, and thus in the definition of $V$. When the
perturbation due to RG running is added, the degeneracy is lifted and a
particular form of $V$ is singled out.  If the initial degeneracy is not
exact, the change of $V$ will be large or not depending on the size of the
perturbation ($\delta_{RG}\Delta m_{ij}^2$) compared with the initial mass
splitting at the scale $M$, $\Delta m_{ij}^2(M)$.

When the RG effect dominates, $V$ evolves quickly from its initial value
$V^{(0)}$ to a stable form (an infrared pseudo-fixed point, IRFP)
$V^{(0)}R_{ij}$ where $R_{ij}$ is a rotation in the plane of the two
quasi-degenerate states $i,j$ such that the perturbation of ${\cal M}_\nu$
is diagonal in the rotated basis (as is familiar from degenerate
perturbation theory). Scenarios in which this takes place are attractive
for two main reasons: {\it 1)} some particular mixing angle will be
selected as a result of $V$ approaching its IRFP form and {\it 2)} the
$i$-$j$ mass-splitting will be essentially
determined at low energy by RG effects.

When $\delta_{RG}\Delta m_{ij}^2\sim \Delta m_{ij}^2(M)$, RG effects can
produce substantial changes in $V$ without getting too close to the IRFP.
This scenario is of interest because, as we show in Sect.~3, the IRFP form
of $V$ in the SM and the MSSM is not in agreement with experimental
observations, while intermediate forms of $V$ can be.

\section{Amplification of mixing angles: SM and MSSM}

As explained above, when two neutrino masses are
quasi-degenerate (and with the  same sign) radiative corrections can have
a large effect on neutrino mixing angles. This offers an 
interesting opportunity for generating large  mixing angles at low
energy as an effect of RG evolution, starting with
a mixing angle that might be small. This possibility has received
a great deal of attention in the literature 
\cite{RG2,CEIN4}, \cite{Tanimoto}--\cite{mohapatra}.
Here we explain why the implementation of this idea in the SM or the 
MSSM is not as appealing as usually believed.
In order to show this we will make much use of the RGEs for physical 
parameters
derived in Ref.~\cite{CEIN4} and collected in Appendix~A for convenience.  

Radiative corrections to $V$ are very small unless
$|\epsilon_\tau\nabla_{ij}\log(M/M_Z)| \simgt 1$ for some $i,j$ 
[see Eqs.~(\ref{RGV}, \ref{Tdef}); here
$\epsilon_\tau\sim y_\tau^2/(16\pi^2)$ and $\nabla_{ij}$ was defined in 
the Introduction] which generically requires mass 
degeneracy,
both in absolute value and sign (i.e. $m_i\simeq 
m_j\Rightarrow|\nabla_{ij}| \gg 1$),
except for the SUSY case with very large $\tan\beta$, and thus large
$\epsilon_\tau$.\footnote{If $|\epsilon_\tau\log(M/M_Z)| \sim
{\cal O} (1)$ the validity of the one-loop approximation is in doubt and
the analysis of RG evolution should be done by numerical integration of
the RGEs in order to capture the leading-log effects at all loops.} In
general, if $m_i\simeq m_j$, so that $|\nabla_{ij}|$ dominates
the RGEs of $V_{\tau i}$, $V_{\tau j}$, these quantities change
appreciably, but the following quantities will be approximately 
constant
\bea
\left.
\begin{array}{l}
\Delta_{ij}V_{\tau i}V_{\tau j}\ ,\vspace{5mm}\\
V_{\alpha l}\ , (l\neq i,j)
\end{array}
\right\} \simeq {\mathrm RG-invariant}\ ,
\label{GenRGinv}
\eea
where $\Delta_{ij}=1/\nabla_{ij}$. The IRFP form
for $V$ can be deduced from Eq.~(\ref{RGV}) and corresponds to $T_{ij}=
0$, which for sizeable $\epsilon_\tau \nabla_{ij}$ implies $V_{\tau
i}= 0$ or $V_{\tau j}= 0$ (depending on the sign of 
$\nabla_{ij}$). Such $V$
can not give the observed angles (it could if the SAMSW solution were
still alive). 

Hence, the IRFP form for $V$ should {\em not} be reached. Still, one may
hope that the RG effects could amplify the atmospheric and/or the solar
angles without reaching the IRFP form of $V$. Such possibility would be
acceptable only if {\it 1)} all mixing angles and mass splittings (which 
are also
affected by the running) agree with experiment and {\it 2)} if this can
be achieved with no fine-tuning of the initial conditions.

We explore in turn the possibility of RG amplification of the
mixing angle in a two-flavour case and then for the 
solar or/and atmospheric angles.

\subsection{The two-flavour approximation}

There are several instances (see below) in which the evolution of a
particular mixing angle is well approximated by a two-flavour model. This
simple setting is very useful to understand the main features of the RG
evolution of mixings and mass splittings, and thus the form of the
infrared fixed points and the potential fine-tuning problems associated
with mixing amplification.

\begin{figure}[t]
\centerline{
\psfig{figure=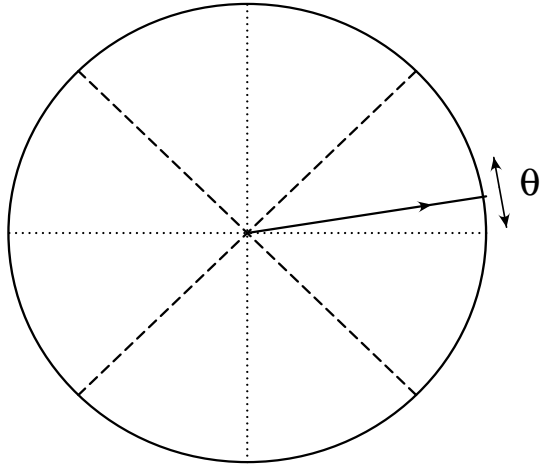,height=6cm,width=5cm,bbllx=4.cm,%
bblly=17.cm,bburx=12.cm,bbury=27.cm}}
\caption{
\noindent{\footnotesize
Pictorial representation of the pseudo-fixed points of the mixing   
angle $\theta$: $0$ and $\pi/2$ in dotted lines, $\pm \pi/4$ in dashed
lines. The initial condition for $\theta$ is represented by a solid line
and the two arrows represent the two possible evolutions of the
running angle.}}
\label{fig:schematic}
\end{figure}
\begin{figure}
\vspace{1.cm}
\centerline{
\vbox{
\psfig{figure=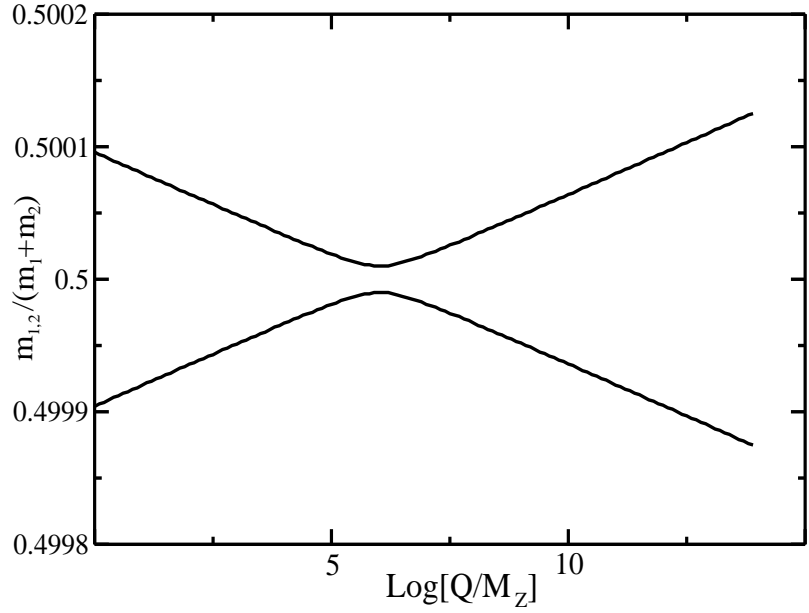,angle=-90,height=7cm,width=7cm,bbllx=11.cm,%
bblly=5.cm,bburx=26.cm,bbury=21.cm}
\psfig{figure=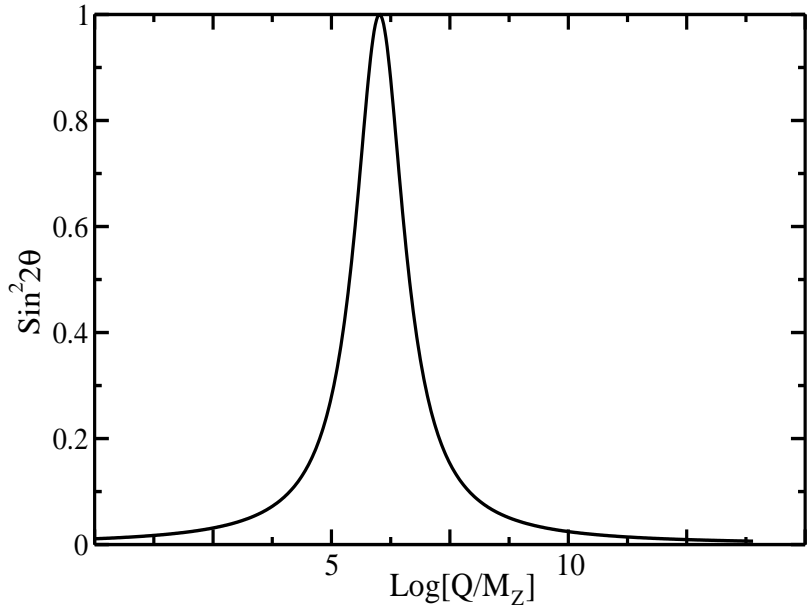,angle=-90,height=7cm,width=7cm,bbllx=5.cm,%
bblly=5.cm,bburx=20.cm,bbury=21.cm}
}}
\caption{
\noindent{\footnotesize Running of $m_1/(m_1+m_2)$ and  $m_2/(m_1+m_2)$ 
(upper plot) and $\sin^22\theta$ (lower plot) from $M$ down to 
$M_Z$ in a two-flavour model width quasi-degenerate masses.}}
\label{fig:2x2}
\end{figure}
In a two-flavour context we have flavour eigenstates, ($\nu_\alpha$,
$\nu_\beta$), a mixing angle, $\theta$, (with $V_{\alpha 2}\equiv 
\sin\theta$ and $\nu_\alpha=\nu_1$ for $\theta=0$) and mass eigenstates
(eigenvalues), $\nu_i$ ($m_i$), $i=1,2$. In a basis where the matrix
of leptonic Yukawa couplings is diagonal, the RGE for the mixing angle
[from Eq.~(\ref{RGV})] takes the form
\bea 
\frac{d \theta}{dt} = -\frac{1}{2}\epsilon_{\alpha\beta}
 \nabla_{21} \sin 2\theta\ .
\label{oldrge}
\eea
with
\be
\epsilon_{\alpha\beta}\equiv c_M\frac{y_\alpha^2-y_\beta^2}{16\pi^2}\ ,
\ee
where $c_M$ is a model-dependent constant.
As previously discussed, for $|\nabla_{21}|\gg 1$
(i.e. for quasi-degenerate neutrinos),
$\theta$ can change appreciably. In such case, it will be
driven towards the infrared (pseudo)-fixed point $\theta^*$ determined
by the condition  $d\theta/dt=0$, which corresponds to 
$\theta^*=0,\pi/2$, that is, towards zero mixing, $\sin
2\theta^*=0$.
The degree of approximation to this fixed point depends on the 
length of
the running interval,  [$\log(M/M_Z)$], on the values of the Yukawa 
couplings, and especially on $\nabla_{21}$.
On the other hand the relative splitting, 
$\Delta_{21}$ satisfies the RGE [from Eq.~(\ref{RGmass})] 
\bea 
\frac{d \Delta_{21}}{dt} = 
\frac{4m_1m_2}{(m_1+m_2)^2}\epsilon_{\alpha\beta}
\cos 2\theta\simeq \epsilon_{\alpha\beta}\cos 2\theta \ ,
\label{RGnabla}
\eea
where, for the last approximation, we have assumed quasi-degenerate neutrinos, 
which is the case of interest. As a consequence, note that 
\be
\Delta_{21}\sin 2\theta\ \simeq\ {\rm RG-invariant}\ .
\label{RGinv}
\ee
There are two qualitatively different possibilities for the running of
$\theta$ depending on the sign of $d\theta/dt$ at $M$ (see
figure~\ref{fig:schematic}, where the fixed points for $\theta$ are indicated
by dotted lines): if $\theta$ decreases with decreasing scale
($d\theta/dt>0$ at $M$) and $\theta^{(0)}\equiv\theta(M)$ is small, the RG 
evolution drives
$\theta$ to zero in the infrared, making it even smaller: the mixing never
gets amplified.  On the opposite case, if $\theta$ increases with
decreasing scale ($d\theta/dt<0$ at $M$), $\theta$ is driven towards
$\theta^*=\pi/2$, and it may happen that the runnig stops (at $M_Z$) near 
$\theta\sim \pi/4$ so that large mixing is obtained.\footnote{
For the solar angle, one should have
$\tan^2\theta_3(M_Z)<1$ (with eigenvalues labelled such that $|m_1|<|m_2|$
holds at low energy), as needed for the MSW solution \cite{MSW}.}
In this second case the RG-evolution is illustrated by 
figure~\ref{fig:2x2}.   The upper plot shows the running of $m_1/(m_1+m_2)$
and $m_2/(m_1+m_2)$ with the scale (this choice removes the universal 
part of the running 
and focuses on the interesting relative mass splitting)  while 
the lower plot shows the running of $\sin^22\theta$. 
Notice that the evolution of the splitting is
quite smooth (first decreasing and then increasing), while the change of
$\theta$ is only important around the scale of maximal mixing ($\theta\sim
\pi/4$) which corresponds to the scale of minimal splitting.  
A simple analytical understanding of this behaviour is possible in the case of 
interest, with quasi-degenerate masses. In that case the RGEs for $\theta$ and 
$\Delta_{21}$, Eqs.~(\ref{oldrge}, \ref{RGnabla}), can be integrated 
exactly 
(assuming also that the running of $y_\tau^2$ is neglected) to get
\bea
\sin^22\theta(Q)&=&\displaystyle{
\frac{
\left[\Delta_{21}^{(0)}\sin2\theta^{(0)}
\right]^2
}{
\left[\Delta_{21}^{(0)}\cos2\theta^{(0)}
-\epsilon_{\alpha\beta}\log{M\over Q}\right]^2
+\left[\Delta_{21}^{(0)}\sin2\theta^{(0)}
\right]^2}
}\ ,\\
\Delta^2_{21}(Q)&=&\left[\Delta_{21}^{(0)}\cos2\theta^{(0)}
-\epsilon_{\alpha\beta}\log{M\over Q}\right]^2
+\left[\Delta_{21}^{(0)}\sin2\theta^{(0)}
\right]^2\ .
\label{exactdelta}
\eea
{}From these solutions we can immediately obtain the scale $Q_{max}$ at which 
maximal mixing occurs:
\be
\log{M\over 
Q_{max}}={\displaystyle{\frac{\Delta_{21}^{(0)}}{
\epsilon_{\alpha\beta}}}}\cos2\theta^{(0)}\ ;
\label{Qmax}
\ee
the half-width, $\omega$, of the 'resonance' (defined at $\sin^22\theta=1/2$)
\be
\omega={\displaystyle{\frac{\Delta_{21}^{(0)}}{
\epsilon_{\alpha\beta}}}}\sin2\theta^{(0)}\ ;
\label{omega}
\ee
and the minimal splitting:
\be
\Delta_{21,min}\equiv\Delta_{21}(Q_{max})=\Delta_{21}^{(0)}\sin2\theta^{(0)}\ 
.
\ee
These results make clear that amplification requires a fine-tuning of 
the initial conditions. Suppose one desires that the initially small 
value of the mixing, $\sin 2\theta^{(0)}$, gets amplified by a factor 
$F\gg 1$ at low energy due to the running. 
From (\ref{RGinv}), this requires the initial relative splitting,
$\Delta_{21}^{(0)}$ to be fine-tuned to the
RG shift, $\delta_{RG}\Delta_{21}$, as
\bea 
\Delta_{21}^{(0)} &=& 
F\Delta_{21}(M_Z) =
F[\Delta_{21}^{(0)} + \delta_{RG}\Delta_{21}]\vspace{5mm}\nonumber\\
& \Rightarrow\
&\delta_{RG}\Delta_{21}=-\left(1-\frac{1}{F}\right)\Delta_{21}^{(0)} \ ,
\label{tuningDelta}
\eea
where\footnote{This is a one-loop leading-log approximation valid when
$\sin^2 2\theta^{(0)}\ll 1$. A more precise result is
given by the exact expression in Eq.~(\ref{exactdelta}), which includes 
all leading-log corrections.}
\bea 
\delta_{RG}\Delta_{21}\simeq
\epsilon_{\alpha\beta}\cos2\theta
\log{M\over M_Z}\ .
\label{deltaRGDelta}
\eea
Hence, Eq.~(\ref{tuningDelta}), which makes  quantitative the arguments of 
the last paragraph of Sect.~2.3, exposes a fine-tuning of one part in $F$ 
between two completely unrelated quantities. 
There is no (known) reason why these two quantities
should be even of a similar order of magnitude, which stresses
the artificiality of such coincidence. 

Alternatively, this fine-tuning
can be seen in the expressions for the scale $Q_{max}$ and the 
half-width $\omega$ [Eqs.~(\ref{Qmax}, \ref{omega})]. The initial 
splitting, $\Delta_{21}^{(0)}$, and the strength of the radiative effect,
$\epsilon_{\alpha\beta}$, have to be right to get $Q_{max}$ near $M_Z$:
If $\Delta_{21}^{(0)}$ is small and/or $\epsilon_{\alpha\beta}$ is 
large, 
the angle 
goes through maximal mixing too quickly; if $\Delta_{21}^{(0)}$ is large 
and/or $\epsilon_{\alpha\beta}$ is small, the angle never grows 
appreciably.
How delicate the balance must be is measured by the 
half-width $\omega$, or better its ratio to the running interval,
\be
\frac{\omega}{\log(M/Q_{max})}=\tan2\theta^{(0)}\ ,
\ee
 which is of order $1/F$ in agreement with the previous estimate of the 
fine-tuning. Finally, notice from (\ref{tuningDelta}) that the 
RG-shift must satisfy $\delta_{RG}\Delta_{21} =
(1-F)\Delta_{21}^{exp}$, which may be 
impossible or unnatural to arrange, as we show in some examples below.

\subsection{Solar angle}

To amplify only the solar angle,
$\theta_{sol}\equiv\theta_{3}$, 
the RGE of $V$ must be dominated by $\nabla_{21}$ [see (\ref{teta3})], 
which
requires a quasi-degenerate ($m_1\simeq m_2\simeq |m_3|$) or
inversely-hierarchical ($m_1\simeq m_2\gg |m_3|$) spectrum. Then the
RG-corrected $V$ is $V^{(0)}R_{12}(\phi)$ with $\phi$ evolving towards
an IRFP such that $V_{\tau 1}= 0$ or $V_{\tau 2}= 0$, while 
$V_{\alpha 3}=(s_2,s_1c_2,c_1c_2)^T$ is almost unaffected by the running. 
This means, in particular,
that $\theta_1$ and $\theta_2$ have to be determined by the physics at
$M$, so as to have $V_{\alpha 3}\simeq (0,1/\sqrt{2},1/\sqrt{2})^T$ as
an initial condition.  An atractive feature of this scenario is that
the running would not upset such initial values as only $\theta_3$ is
affected. This is most clearly seen by realizing that $V\rightarrow
V^{(0)}R_{12}(\phi)$ amounts simply to $\theta_3\rightarrow
\theta_3^{(0)}+\phi$ [see (\ref{Vdef})].

It is a good approximation to treat the running of $\theta_3$ in a
two-flavour context [see Eqs.~(\ref{oldrge}) and (\ref{teta3})] with
$(\nu_\alpha,\nu_\beta)\equiv (\nu_e,\
\nu_\mu\cos\theta_{1}-\nu_\tau\sin\theta_{1})$, and 
$(m_i,m_j)=(m_1,m_2)$.  Therefore we can apply the results
obtained in the previous subsection [in particular 
Eqs.~(\ref{oldrge}, \ref{RGnabla})
with $y_\alpha^2\simeq 0$, $y_\beta^2\simeq s_1^2 y_\tau^2\simeq
y_\tau^2/2$] to conclude that the amplification of $\sin 2 
\theta_3$
by a factor $F\gg 1$ requires a fine-tuning of one part in $F$ between
two completely unrelated quantities: the relative mass splitting at
the $M$ scale, $\Delta_{21}^{(0)}$, and the splitting  generated
radiatively, $\delta_{RG}\Delta_{21}$.

Moreover, from Eqs.~(\ref{tuningDelta}, \ref{deltaRGDelta}),
$\delta_{RG}\Delta_{21}\simeq (\epsilon_\tau/2) \log(M/M_Z)$
has to match $(1-F)\Delta_{21}^{exp}$. In the SM 
$\delta_{RG}\Delta_{21}$
is quite small, which means that $F$ must be close to one:
$F-1 \simeq  10^{-3} \log(M/M_Z)({7\times  10^{-5} {\rm eV}^2
/\Delta m^2_{sol}})({m/ 0.2\  {\rm eV}})^2$.
Consequently, it is
not possible to amplify the solar angle in the SM, even with
fine-tunings.
Larger values of $ \delta_{RG}\Delta_{21}$ can be
achieved in the MSSM for large $\tan \beta$: amplification of $\sin 2
\theta_3$ by a factor $F$ requires
\be
\tan\beta\simeq
37\sqrt{{(F-1)\over\log(M/M_Z)}}\left({\Delta m^2_{sol}\over 7\times  
10^{-5}\  {\rm eV}^2}\right)^{1/2}
\left({0.2\  {\rm eV}\over m}\right)\ ,
\ee
where $\log(M/M_Z)\sim 30$ is a typical value.

\subsection{Atmospheric angle}

This case was critically examined already in \cite{CEIN4,nogomax}. We 
summarize here the main arguments and results and complete the analysis.  From
Eqs.~(\ref{teta1}--\ref{teta3}), the amplification of
$\theta_{atm}\equiv\theta_{1}$ requires that $\nabla_{31}$ or
$\nabla_{32}$ dominate the RGEs, and therefore $m_1\simeq -m_2\simeq \pm
m_3$ is necessary.\footnote{Another possibility, with $m_1\simeq m_2\simeq
m_3$, is discussed in Sect.~3.4.}
 
\begin{figure}[tp]
\centerline{
\psfig{figure=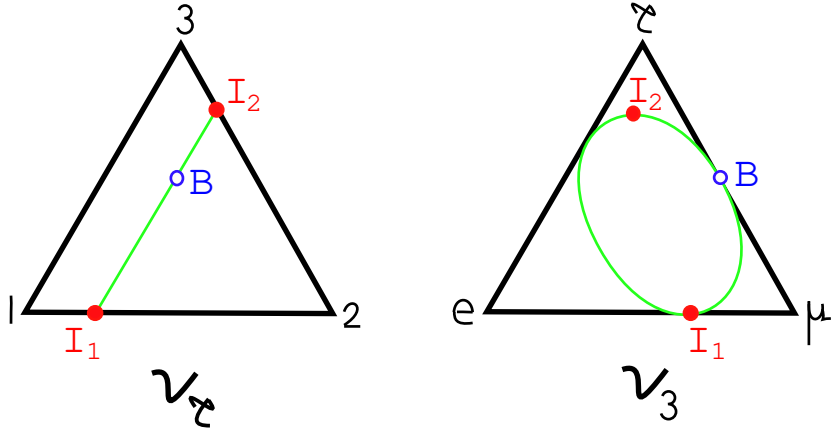,height=5.7cm,width=5cm,bbllx=7.cm,%
bblly=18.cm,bburx=14.cm,bbury=26.cm}}
\caption{
\noindent{\footnotesize RG trajectory of $V$ in matrix space for
$|\nabla_{32}|\gg 1$. The red solid dots represent IRFPs, while the blue 
empty dots represent the bi-maximal mixing point for the atmosferic 
and solar angles.
}}
\label{fig:godseye}
\end{figure}

Suppose that $\nabla_{31}$ is dominant (for $\nabla_{32}$ the argument
is similar). The RG-corrected mixing matrix takes
the form $V=V^{(0)}R_{31}(\phi)$, with $\phi$ evolving towards an IRFP
such that $V_{\tau 1}= 0$ or $V_{\tau 3}= 0$ while $V_{\alpha 
2}$
remains almost constant.\footnote{The two-flavour approximation  is not 
possible here: it 
requires $s_2\simeq s_3\simeq 0$, at odds with experiment.}   Therefore, 
to agree with
experiment we must assume as an initial condition $V_{\alpha 2}\simeq
(1/\sqrt{2},1/2,-1/2)^T$. As $\phi$ changes, the path of $V$ in 
matrix space goes through a bi-maximal mixing form. This (closed) path is 
represented in figure~\ref{fig:godseye} which makes use of triangular 
diagrams \cite{triangles} to represent $V$ by a pair of points inside two 
equilateral triangles of unit height. A point inside the left triangle 
(say $B$)
determines three distances to each side, which correspond to 
$|V_{\tau 1}|^2$, $|V_{\tau 2}|^2$ and $|V_{\tau 3}|^2$ (with the unitary 
condition $\sum_i|V_{\tau i}|^2=1$ ensured geometrically). On the right 
triangle the point $B$ determines instead  
$|V_{e 3}|^2$, $|V_{\mu 3}|^2$ and $|V_{\tau 3}|^2$ (redundance in the 
latter requires the points to be drawn with the same height on 
both sides). 

In figure~\ref{fig:godseye} the green path is traversed as 
$\phi$ is varied, with IRFPs marked by solid red dots ($I_1$, $I_2$).
 By assumption  
we start the running at some point in this path near $\sin^22\theta_1=0$,
{\it i.e.} near the intersections with the $e\tau$-side ($\theta_1=0$)
or the $e\mu$-side ($\theta_1=\pi/2$). Maximal $\theta_1$ mixing 
corresponds to points equidistant from these two sides. The goal would be
to stop the running near the point $B$ marked by the open blue dot, 
which 
has 
maximal $\theta_1$ mixing and zero $\theta_2$. From the location of the 
IRFP points, this amplification can only work if we start near 
$\theta_1=\pi/2$ (starting near $\theta_1=0$ we can never reach our goal)
and have $|V_{e3}|^2<1/3$ (so as to take the right path). 
Schematically, this right path from one IRFP to the other going through 
the point of bi-maximal mixing is
\bea
V_{I_1}\equiv\pmatrix{\frac{1}{\sqrt{6}} &  \frac{1}{\sqrt{2}} &
-\frac{1}{\sqrt{3}} \cr \frac{1}{2\sqrt{3}}   &  \frac{1}{2} &
\frac{2}{\sqrt{6}} \cr \frac{\sqrt{3}}{2}  & -\frac{1}{2}  & 0 \cr}
\
\rightarrow  
V_{B}\equiv
\pmatrix{\frac{1}{\sqrt{2}} & \frac{1}{\sqrt{2}}  &   0
\cr -\frac{1}{2} &  \frac{1}{2}  &  \frac{1}{\sqrt{2}} \cr \frac{1}{2}
& -\frac{1}{2} & \frac{1}{\sqrt{2}} \cr}
\ 
\rightarrow
V_{I_2}\equiv
\pmatrix{-\frac{1}{\sqrt{3}} & \frac{1}{\sqrt{2}}   &
\frac{1}{\sqrt{6}} \cr \frac{2}{\sqrt{6}} &   \frac{1}{2}  &
\frac{1}{2\sqrt{3}} \cr 0 & -\frac{1}{2}  &  \frac{\sqrt{3}}{2}\cr}
\nonumber
\eea
(a change in the sign of $\nabla_{31}$ reverses the direction of the
arrows). Under these assumptions amplification at the right scale still 
requires a certain fine-tuning, similar to the one  dicussed in the
previous section. 

In addition there is now an even worse drawback
because it is difficult to do this tuning without upsetting $\Delta
m^2_{sol}$: the relative mass splittings are approximately given by
[from (\ref{RGmass})]
\bea
\label{atmsplit}
\frac{\Delta m^2_{atm}}{m^2}&=&  \left[\frac{\Delta
m^2_{31}}{m^2}\right]^{(0)} +4\epsilon_{\tau}\langle V_{\tau
3}^2-V_{\tau 1}^2\rangle \log{M\over M_Z}\ ,\\ \frac{\Delta
m^2_{sol}}{m^2}&=&  \left[\frac{\Delta m^2_{21}}{m^2}\right]^{(0)}
+4\epsilon_{\tau}\langle V_{\tau 2}^2-V_{\tau 1}^2\rangle\log{M\over
M_Z}\ ,
\label{solsplit}    
\eea where $\langle\rangle$ denote averages over the interval of
running and $m$ is the overall neutrino mass. We see that both
radiative corrections are of similar magnitude because, for rapidly
changing $V_{\tau 1}$ and $V_{\tau 3}$, the averages of matrix
elements in (\ref{atmsplit}, \ref{solsplit}) cannot be suppressed and
are therefore of ${\cal O}(1)$.   On the other hand,
$\delta_{RG}(\Delta m^2_{31}/m^2)$ [last term in Eq.~(\ref{atmsplit})] 
must be
$(F-1)\Delta m^2_{atm}/m^2$, where $F$ is defined as $V_{\tau
1}V_{\tau 3}= F V_{\tau 1}^{(0)}V_{\tau 3}^{(0)}$ 
[see Eq.~(\ref{GenRGinv})]. For a sizeable amplification, $(F-1)\simgt 1$, 
so unless there is
an extremely accurate and artifitial cancellation in (\ref{solsplit}),
one naturally expects $\Delta m^2_{sol} \sim \Delta m^2_{atm}$, which
is not acceptable. In fact, there are further problems: in the SM
$\delta_{RG}(\Delta m^2_{31}/m^2)$ is not large enough to match
$(F-1)\Delta m^2_{atm}/m^2$. In the MSSM this is possible, but it
requires, besides a certain tuning, a very large $\tan \beta$ ($\simgt
100$ for $|m|<0.3$ eV).

\subsection{Solar and atmospheric angles simultaneously}

There is still a possibility for mixing amplification not discussed
in the previous subsections (3.2 and 3.3),
namely when $m_1\simeq m_2\simeq m_3$, both in absolute value and
sign\footnote{This possibility has been recently used in \cite{mohapatra},
where the implicit fine-tuning we are about to show was not addressed.}. 
Then $\nabla_{21}\gg \nabla_{31}\simeq \nabla_{32}\gg 1$ (in absolute
value). 
Notice from Eq.~(\ref{teta1}--\ref{teta3}) that if $\epsilon_\tau
\nabla_{31}\log(M/M_Z) \sim {\cal O}(1)$, so that the atmospheric angle
can run appreciably, then it is mandatory that $V_{\tau 1}V_{\tau 2}\simeq
0$; otherwise the running of $V$ will be strongly dominated by the term
proportional to $T_{21}=\epsilon_\tau V_{\tau 1}V_{\tau 2}\nabla_{21}$, 
and
thus rapidly driven to an IRFP (a phenomenological disaster).
To avoid this, the condition $V_{\tau 1}V_{\tau 2}\simeq 0$ must be
fulfilled initially and along most of the running.
\footnote{This condition implies that one starts near one of the IRFPs.
To avoid falling towards it, the signs of the splittings must 
be such that $V$ is eventually driven towards a different IRFP,  
crossing in its way regions of parameter space with sizeable  $V_{\tau 
1}$ and $V_{\tau 2}$. [This is similar to the situation discussed after 
Eq.~(\ref{RGinv})]}
 In addition one has to demand $V_{\tau 2}^2-V_{\tau 1}^2\simeq 0$ along
most of the running, otherwise $\Delta m^2_{21}$ gets radiative
corrections of a size similar to $\Delta m^2_{atm}$ for the reasons
explained in Subsect.~3.3.  In consequence the possibility under
consideration can only work if $V_{\tau 2}\simeq V_{\tau 1}\simeq 0$
$\Rightarrow$ $\sin \theta_1\simeq \sin \theta_2\simeq 0$ along most of
the running.

The previous conclusion implies that $\theta_1$ must be radiatively amplified 
at low energy. Actually, it can be checked from 
Eq.~(\ref{teta1}) that,
since $\nabla_{31}\simeq \nabla_{32}$, the running of $\theta_1$ is
well approximated by a two-flavour equation (see Subsect.~3.1)
\bea 
\frac{d \theta_1}{dt} \simeq 
\frac{1}{2} \epsilon_\tau \nabla_{31} \,\sin 2\theta_1 \ .
\label{teta1sim}
\eea
Hence, the results of Subsect.~3.1 apply here and we conclude
that the amplification of $\theta_1$ requires {\it 1)} a very large
$\tan \beta$ ($\simgt
100$ for $|m|<0.3$ eV) to get a large enough 
$\delta_{RG} \Delta_{31}$ and {\it 2)} 
a fine-tuning
between the initial splitting $\Delta_{31}^{(0)}$ and the radiative 
correction $\delta_{RG} \Delta_{31}$:
\be
\delta_{RG} \Delta_{31}=-\left(1-{1\over F}\right)\Delta_{31}^{(0)}\ .
\ee
This tuning can also be seen by equations similar to (\ref{Qmax}) and 
(\ref{omega}) which in this 
particular case read
\be
\log\frac{M}{M_Z}\simeq \frac{\Delta_{31}^{(0)}}{
\epsilon_\tau}\cos2\theta_1^{(0)}\ ,\,\,\,
\omega_1\simeq \frac{\Delta_{31}^{(0)}}{
\epsilon_\tau}\sin2\theta_1^{(0)}\ .
\ee
\begin{figure}[t]
\vspace{1.cm}
\centerline{
\psfig{figure=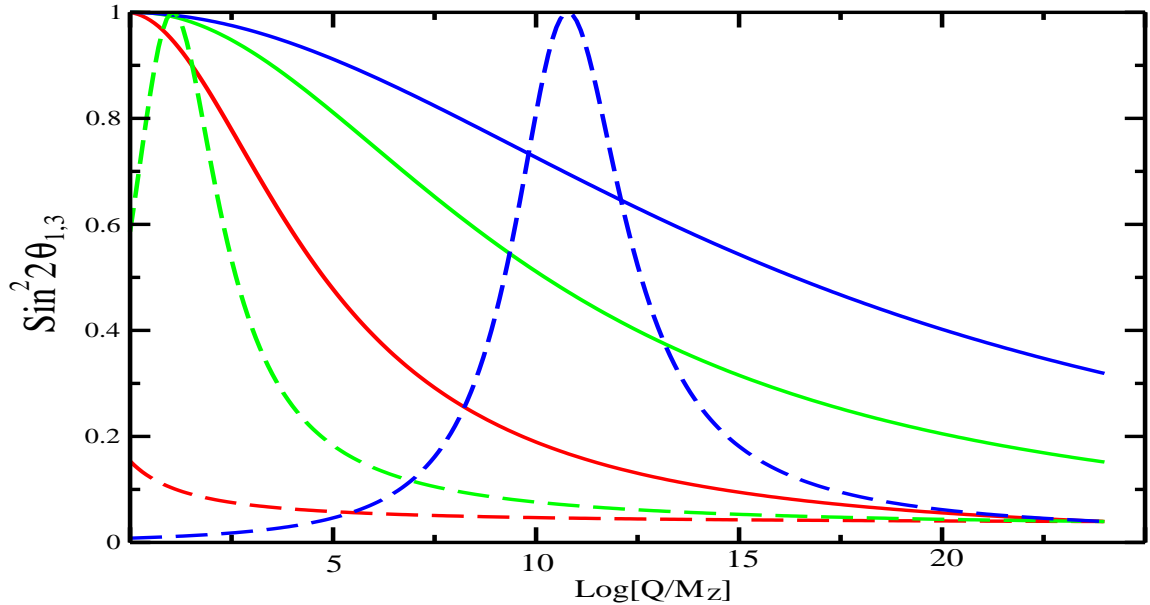,angle=-90,height=7cm,width=10cm,bbllx=5.cm,%
bblly=5.cm,bburx=20.cm,bbury=21.cm}}
\caption{
\noindent{\footnotesize Running of $\sin^22\theta_1$ (solid lines) and 
$\sin^22\theta_3$ (dashed lines)
from $M$ down to $M_Z$ for different initial values of 
$\sin^22\theta_1$.}}
\label{fig:3x3}
\end{figure}
Besides this, another problem affects the running of $\theta_3$. 
The RGE of $\theta_3$ is given by [see Eq.~(\ref{teta3})]
\be 
\frac{d \theta_3}{dt} \simeq 
\frac{1}{2}  (\epsilon_\tau \sin^2\theta_1)
\nabla_{21} \,\sin 2 \theta_3 \ ,
\label{teta3sim}
\ee
which is very similar to a two-flavour RGE except for the extra factor 
$\sin^2\theta_1$.
Clearly, $\sin 2 \theta_3$ must be initially small. Otherwise, since
$|\nabla_{21}|\gg |\nabla_{31}|$, for moderate values of 
$\sin^2\theta_1$, $\theta_3$ runs much more quickly than 
$\theta_1$ and therefore it is driven to the IRFP (with small $\sin 2 
\theta_3$) before $\theta_1$ gets properly amplified.
Consequently, $\theta_3$ needs radiative amplification and this requires 
its own fine-tuning:
\be
\label{Qmax3}
\log\frac{M}{Q_{3,max}}\simeq \frac{\Delta_{21}^{(0)}}{
\epsilon_\tau\sin^2\theta_1^{(0)}}\cos2\theta_3^{(0)}\ ,\,\,\,
\omega_3\simeq \frac{\Delta_{21}^{(0)}}{
\epsilon_\tau\sin^2\theta_1(Q_{3,max})}\sin2\theta_3^{(0)}\ .
\ee
Notice that for estimating the position of the peak in the running of 
$\sin^22\theta_3$
we have simply used the initial value of $\sin^2\theta_1$ as if it would 
not run, while for estimating the half-width, $\omega_3$, a better choice 
is 
to use $\sin^2\theta_1$ at the peak $Q_{3,max}$, which is assumed to be 
around $M_Z$.
This means in particular that $\omega_3$ is not enlarged 
significantly by the initiall smallness of $\sin^2\theta_1$ and, being 
controlled by 
$\Delta_{21}^{(0)}$, it is in general even smaller than $\omega_1$.
This behaviour is shown in figure \ref{fig:3x3} where the solid lines give
$\sin^22\theta_1$ and the dashed ones $\sin^22\theta_3$. The three 
different pairs of curves correspond to different initial conditions for
$\sin^22\theta_1$, with the $\Delta_{31}^{(0)}$ mass splitting chosen so 
as to 
get maximal atmospheric mixing at $M_Z$. As 
expected from Eq.~(\ref{Qmax3}), when $\sin^22\theta_1^{(0)}$ 
decreases, the narrow peak in $\sin^22\theta_3$ moves rapidly to lower 
scales, making clear the need for an extra fine-tuning to ensure a large 
solar mixing angle at $M_Z$.

\section{Amplification of mixing angles: K\"ahler masses}

Let us consider now the possibility of radiative amplification of mixing
angles in the scenarios described in Section~2.2, {\it i.e.} when neutrino
masses in a supersymmetric model originate from non-renormalizable
operators in the K\"ahler potential \cite{PRL}.  More precisely, we
consider only the operator $\kappa (L\cdot H_2)(L\cdot\overline{H}_1)$ as
discussed in Sect.~2.2. Then the RG-evolution of mixing angles is
described by an equation of the usual form:
\be
 {dV\over dt}=VT\ ,
\ee
with the matrix $T$ given by (see Appendix~A):
\be
T_{ij}=
{-1\over 16\pi^2}(y_\alpha^2-y_\beta^2){V_{\alpha i}V_{\beta j}\over
m_i-m_j}m^A_{\alpha\beta}\ ,
\label{Tij}
\ee
for $i\neq j$ and $T_{ii}=0$. Here 
$m^A_{\alpha\beta}\equiv(M^A_\nu)_{\alpha\beta}$, where the matrix 
$M^A_\nu$ is related to 
the K\"ahler matrix $\kappa_A$ by $M_\nu^A\equiv\mu\kappa_A v^2/M^2$. The 
neutrino 
mass eigenvalues run according to (no sum in $i$)
\be
\label{dmi}
\frac{d m_i}{dt}= u m_i
+ {y_\alpha^2\over 8\pi^2} V_{\alpha i}m^A_{\alpha\beta}V_{\beta i} \ .
\ee
The generic condition required to have a significant change in the mixing 
angles is to have some sizeable $T_{ij}$, i.e. 
\be
{y_\tau^2\over 16\pi^2}
{m^A_{\alpha\tau} \over m_i-m_j}\log{M\over M_Z}
\simgt
{\cal O}(1)\ .
\label{newcond}
\ee
This is consistent with the general arguments of Subsect.~2.3: 
notice from (\ref{dmi}) that the relative splitting,
$\Delta_{ij}$, typically gets a correction
\be
\delta_{RG}\Delta_{ij} \sim {\cal O}\left(
{y_\tau^2\over 16\pi^2}{m^A_{\alpha\tau} \over m_i+m_j}
\log{M\over M_Z}
\right)
\label{deltaDelta}\ .
\ee
Therefore, important effects in the mixing angles,
Eq.~(\ref{newcond}), occur when the RG-induced (relative) splittings
are comparable (or larger) than the initial splitting 
($\delta_{RG}\Delta_{ij} \simeq \Delta_{ij} $).

Comparing Eqs.~(\ref{Tij}, \ref{newcond}) with the 
conventional  $T_{ij}$, Eq~(\ref{Tdef}),
we see that it is possible to have now large effects even for
neutrino spectra without quasi-degenerate masses, provided the magnitude 
of the entries of $M^A_\nu$ is larger than the mass differences 
$m_i-m_j$.
This implies that, in the new 
scenario, amplification of mixing angles is a more general phenomenon, 
which can occur also for spectra that cannot accomodate amplification in 
conventional models, {\it e.g.} normal hierarchy or inverted hierarchy 
with $m_1\simeq - m_2$. 

In parallel with the discussion of the conventional case (Sect.~3) we 
consider 
in turn the amplification for a  two-flavour case, for the solar angle and 
for the atmospheric angle.

\subsection{Two-flavour scenario}

\begin{figure}[t]
\vspace{1.cm}
\centerline{
\psfig{figure=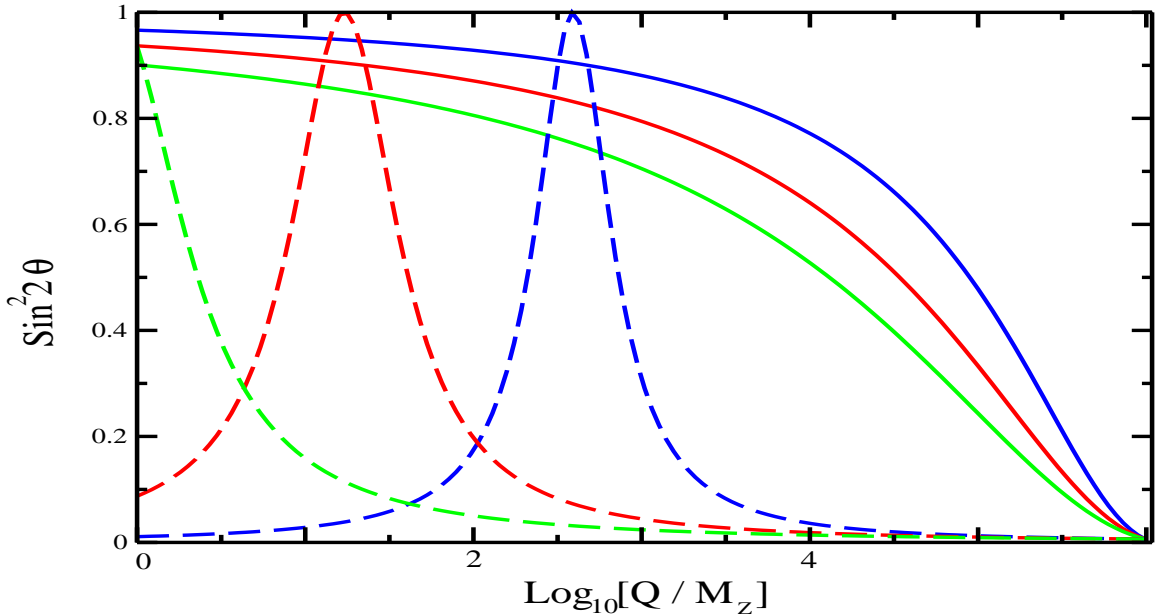,angle=-90,height=7cm,width=10cm,bbllx=5.cm,%
bblly=5.cm,bburx=20.cm,bbury=21.cm}}
\caption{
\noindent{\footnotesize Running of $\sin^22\theta$ from $M$ down to 
$M_Z$ in conventional scenarios (dashed lines) as compared with 
the K\"ahler case (solid lines). In each case, different lines 
correspond to different mass splittings at $M$.}}
\label{fig:amplif}
\end{figure}

As for the conventional case, the two-flavour model  is very useful to
understand in a simple setting the main features of the RG evolution
of mixings and mass splittings, and the form of the infrared fixed
points.
In this new scenario the evolution of the mixing 
angle in a two flavour  case does not follow an RGE 
of the form (\ref{oldrge}) but rather the following (no sum in 
$\alpha,\beta$):
\bea 
\frac{d \theta}{dt} = -\epsilon_{\alpha\beta}
\frac{m^A_{\alpha\beta}}{m_1 - m_2}
\cos 2\theta 
\ ,
\label{newrge}
\eea
where again  $\epsilon_{\alpha\beta}\equiv
(y_\alpha^2-y_\beta^2)/(16\pi^2)$ with $\alpha$ and $\beta$ the
two flavours and $V_{\alpha 2}\equiv \sin\theta$.  Comparing 
Eq~(\ref{newrge}) to Eq~(\ref{oldrge})
(i.e. the one for the  conventional SM and MSSM cases) we note, beside
numerical factors,  two main differences: the quantity
$\nabla_{12}=(m_1+m_2)/(m_1-m_2)$ is replaced by the ratio
$m^A_{\alpha\beta}/(m_1-m_2)$, and $\sin 2\theta$ by $\cos
2\theta$. The first difference implies that important changes on $\theta$ 
are
achieved for $|m^A_{\alpha\beta}|\gg |m_1-m_2|$, which does not require
quasi-degeneracy of the eigenvalues. The second  one implies that the
running will drive $\cos 2\theta$ (rather than $\sin 2\theta$) to zero,
i.e. IRFPs for $\theta$ are  now at maximal mixing, $\theta^*=\pm\pi/4$  
(dashed lines in
fig.~\ref{fig:schematic}). Therefore,  $\theta$ will be amplified
towards maximal mixing  in the infrared\footnote{Again,
for the solar angle, one should have
$\tan^2\theta_3(M_Z)<1$ (with eigenvalues labelled such that $|m_1|<|m_2|$
holds at low energy), as needed for 
the MSW solution \cite{MSW}.} 
irrespective of the sign of
$d\theta/dt$ at $M$ and  with no fine-tuning required on $m_1-m_2$.

For comparison with the
conventional case, we have plotted 
in fig.~\ref{fig:amplif} the new evolution of $\sin^22\theta(Q)$ in solid 
lines, for different initial mass splittings; the dashed 
lines correspond to the running in the conventional scenario (with the 
same initial conditions). Clearly, 
in order to generate large neutrino mixing angles through RG evolution, 
scenarios that follow Eq.~(\ref{newrge}) are more natural than those that 
follow Eq.~(\ref{oldrge}).
Let us mention that, since $M$ is smaller now,
the interval in energy available to amplify $\theta$ is also smaller.

Regarding the mass splittings, the quantity 
$(m_1-m_2)/m^A_{\alpha\beta}$, which is now the relevant one, 
satisfies the RGE  [from Eqs.~(\ref{kappaS}) and (\ref{kappaA})]
\bea 
\frac{d}{dt}\left[{m_1-m_2\over m^A_{\alpha\beta}}\right] 
&=& 
-2\epsilon_{\alpha\beta}\sin 2\theta\left[1-{3\over 4}\left( 
{m_1-m_2\over m^A_{\alpha\beta}}\right)^2\right]
-{1\over 8\pi^2}(y_\alpha^2+ y_\beta^2)
{m_1-m_2\over m^A_{\alpha\beta}}\nonumber\vspace{5mm}\\
&\simeq &
-2\epsilon_{\alpha\beta}\sin 2\theta \ ,
\label{RGnablaK}
\eea
where, for the last approximation, we have assumed 
$|m^A_{\alpha\beta}|\gg |m_1-m_2|$, which is the case of interest.
As a consequence, note that 
\be
{m_1-m_2\over m^A_{\alpha\beta}}\cos 2\theta\ \simeq\ {\rm RG-invariant}\ 
,
\label{RGinvK}
\ee
in contrast with Eq.~(\ref{RGinv}).

\subsection{Solar angle}

In the new scenario many of the requirements for amplification of the 
solar angle are the same as those in conventional cases: the RGE of $V$
should be dominated by $T_{21}$; $V_{\alpha 3}$ does not run 
appreciably and should be fixed as an initial condition to be 
$\simeq (0,1/\sqrt{2},1/\sqrt{2})^T$ while $\theta_3$ is the only angle 
that changes significantly. The difference with respect to the standard
case is that $\theta_3$ is now driven towards a different IRFP, 
determined by the condition $T_{21}\rightarrow 0$. In terms of $\theta_3$ 
and neglecting all leptonic Yukawa couplings other than $y_\tau$ this 
IRFP condition reads
\be
\label{IRFPsolar}
\tan 2\theta_3\rightarrow {2\sin\theta_2\cos 2\theta_1 
m^A_{\mu\tau}
+2\sin\theta_1\cos \theta_2 m^A_{e\tau}
\over
\sin 2\theta_1(1+\sin^2\theta_2) m^A_{\mu\tau}
-\cos\theta_1\sin 2\theta_2 m^A_{e\tau}}\ .
\ee
As this equation clearly shows, the IR behaviour of the running $\theta_3$
does not correspond in general to that expected in a two flavour case 
(discussed in the previous subsection) but is richer. Several cases of 
interest are the following:
\begin{itemize}
\item If we make the approximation $\sin\theta_2\simeq 0$, $\theta_3$
evolves towards the IRFP 
$\tan 2\theta_3^*\simeq m^A_{e\tau}/[\cos\theta_1 
m^A_{\mu\tau}]$ and, for $m^A_{e\tau}/m^A_{\mu\tau}\sim 
{\cal O}(1)$ one can easily get $\theta_3^*$ inside the experimental 
range.

\item If, in the previous case with $\sin\theta_2\simeq 0$ one has 
$m^A_{e\tau}\gg m^A_{\mu\tau}$ 
(the precise condition is 
$s_{2\theta_2}\ll m^A_{\mu\tau}/m^A_{e\tau}\ll 1$) 
the IRFP for $\theta_3$ is at maximal mixing. This 
case is indeed well described by a two-flavour approximation and can be 
acceptable if the running interval is not too long so that the low-energy 
value of $\theta_3$ is not too close to the IRFP.

\item If $m^A_{\mu\tau}\simeq 0$ [or, more precisely, 
$m^A_{\mu\tau}\ll s_{2\theta_2}m^A_{e\tau}$], the IRFP is 
simply $\tan 2\theta_3^*\simeq -\tan\theta_1/\sin\theta_2$. That is, this 
scenario predicts 
a correlation between the neutrino mixing angles such that, given the 
experimental interval for $\theta_3$, the angle $\theta_2$ is predicted to 
be in the range $0.02 \simlt \sin^2\theta_2\simlt 0.5$ (for 
$\tan\theta_1=1$). In other words,
for $\theta_3^*$ in the upper region of its experimentally
allowed range, $\theta_2$ lies not far below the CHOOZ bound.
\end{itemize}

To end the discussion of the solar case one should also check that the
requirement of phenomenological low-energy mass splittings 
is not in conflict with the requirements just described needed to
amplify the solar angle.  The solar mass splitting, in one-loop
leading-log approximation\footnote{Two-loop leading-log corrections $\sim
[y_\tau^2 m^A_{\alpha\tau}\log(M/M_Z)/(16\pi^2)]^2$ produce a
non-vanishing splitting at low energy even if initially $m_i=0$. Such
corrections can be easily obtained from Eq.~(\ref{dmi}).}, is now given by
\bea
\Delta m_{sol}^2 &\simeq& 
\Delta m_{21}^2(M)\left[1-2u \log{M\over M_Z}\right] \\
&+&{y_\tau^2\over 4\pi^2}
\left[(V_{e2}V_{\tau 2}m_2-V_{e1}V_{\tau 1}m_1)m^A_{e\tau}+
(V_{\mu 2}V_{\tau 2}m_2-V_{\mu 1}V_{\tau 1}m_1)m^A_{\mu\tau}
\right]\log{M\over M_Z}\ .
\nonumber
\eea
If the running is long enough to approach the IRFP, in which case 
the 
radiative correction of the splitting overcomes the initial
value at the $M$--scale [see the discussion around Eq.~(\ref{newcond})],
the above result gets simplified to
\be
\Delta m_{sol}^2 \simeq (m_1+m_2)(M_Z)
{y_\tau^2\over 8\pi^2}\left[(
V_{e2}V_{\tau 2}-V_{e1}V_{\tau 1})m^A_{e\tau}+
(V_{\mu 2}V_{\tau 2}-V_{\mu 1}V_{\tau 1})m^A_{\mu\tau}
\right]\log{M\over M_Z}\ .
\label{DeltasolK}
\ee
The most important aspect of this formula is that, in stark contrast to 
the standard cases of the SM or the MSSM, the radiative correction to the 
mass splitting is controlled by the elements of the matrix $M_\nu^A$, 
which do not contribute at tree level to neutrino masses. It is therefore
quite easy to arrange the magnitude of the mass splitting at will, and 
this without disturbing the IRFP, which does not depend on the 
overall magnitude of $M_\nu^A$ but only on the ratio of its elements
[see Eq.~(\ref{IRFPsolar})].
Note also that the global scale of $M^A_\nu$, which is in principle
an unknown in this kind of scenarios, is thus fixed to get
the correct mass splitting. In conventional scenarios the latter
is adjusted by tuning the initial mass splitting at high energy, as
discussed in the previous section.

\subsection{Atmospheric angle}

As in the conventional cases, amplification of the atmospheric angle
requires that $T_{31}$ or $T_{32}$ dominate the RGEs of the  neutrino
mixing angles [we will not discuss the case $T_{31}\simeq
T_{32}$ which in this scenario seems an unnatural coincidence in view
of Eq.(\ref{Tij})].  Both cases are in fact very similar, so we
consider in some detail only the case of $T_{31}$-dominance and later
explain what changes would  apply to the other case.  Incidentally,
notice from (\ref{Tij}) that a natural way to get $T_{31}\gg
T_{21}$ is $|m_2-m_1|\gg |m_3-m_1|$, implying $m_1\simeq -m_2\simeq
m_3$, as for the conventional case.

If $T_{31}$ dominates the evolution of the angles, the column
$V_{\alpha 2}$ does not change much and should be chosen to agree with
experimental data. This is imposed by hand as an initial condition, to
be explained by physics at higher energy scales. By unitarity, this
requirement amounts to only two conditions on the mixing angles. The
IRFP is determined as usual by the condition $T_{31}\rightarrow 0$,
which leads to the third condition on the mixing angles: \be
\label{IRFPatm31}
(c_1 c_3 c_{2\theta_2}+s_1s_2s_3)m^A_{e\tau} \simeq
c_2(s_3c_{2\theta_1}+s_2s_{2\theta_1}c_3) m^A_{\mu\tau}\ .  \ee This
IRFP condition amounts to one prediction for the angles in these
scenarios. 

One interesting possibility for the IRFP is the following. If
$m^A_{e\tau}\ll m^A_{\mu\tau}$ the IRFP (\ref{IRFPatm31}) reads $\tan
2\theta_1\rightarrow -s_3/(s_2c_3)$. This correlation between angles
relates the smallness of the CHOOZ angle to the maximality of the
atmospheric one and can be taken as an interesting prediction of this
particular scenario.

The case of $T_{32}$-dominance is obtained from the previous one by
the replacement $c_3\rightarrow s_3$ and $s_3\rightarrow -c_3$ (which
interchanges the first and second columns of $V$). The IRFP condition
is  then 
\be
\label{IRFPatm32}
(c_1 s_3 c_{2\theta_2}+s_1s_2c_3)m^A_{e\tau} \simeq
c_2(-c_3c_{2\theta_1}+s_2s_{2\theta_1}s_3) m^A_{\mu\tau}\ , \ee and
all implications that follow are quite similar to the ones discussed
above.

As for the solar case, provided the running reaches the IRFP,
the low-energy mass splitting  $\Delta m_{13}^2$ is of purely radiative
origin and given by an expression similar to
(\ref{DeltasolK}). Again the global magnitude of $M_\nu^A$ can be
chosen to reproduce the atmospheric splitting, without modifying the
previous discussion of IRFPs for the atmospheric angle, which are
just  controlled by the ratios  of $M_\nu^A$ entries [see
Eqs.~(\ref{IRFPatm31}, \ref{IRFPatm32})].  Furthermore, in contrast
with the conventional cases discussed in Sect.~3.3, this is achieved
without upsetting the solar mass splitting. To see this notice that
$m_1\simeq -m_2\simeq \pm m_3$ implies that the radiative correction
to the solar splitting, still given by Eq.(\ref{DeltasolK}), is now of
the order $\delta_{RG}\Delta m^2_{sol}\sim \Delta m_{atm}^2\Delta 
m^2_{sol}/m^2$.

\section{Conclusions}

The nearly bi-maximal structure of the neutrino mixing matrix, $V_{PMNS}$,
is very different from that of $V_{CKM}$ in the quark sector, where all
the mixings are small. An attractive possibility to understand this is
that some neutrino mixings are radiatively enhanced, i.e. are initially
small and get large in the RG running from high to low energy.

In this paper we have carefully examined this issue in two different
contexts: conventional scenarios, in particular the SM and the MSSM,
and supersymmetric scenarios in which neutrino
masses originate in the K\"ahler  potential. The RGEs are quite
different in each case, and so are the results and conclusions. Our
analysis is complete in the sense that we have taken into account all
neutrino parameters, to ensure that not only mixing angles but also
mass splittings agree with experiment at low energy. Moreover we have
investigated which scenarios require a fine-tuning in order to achieve
the amplification, and quantified it. In order to explain in an
intuitive way the main issues involved in the running
(appearance of infrared fixed points, interplay between
the running of the angles and that of the mass splittings, origin of
the fine-tunings, etc.) we have first studied the two-flavour
scenario, where all these features show up in a transparent way.
Then, we have explored the physical cases exhaustively.
Our main results and conclusions are the following.

\begin{itemize}

\item In the SM it is not possible to amplify either the solar or the
atmospheric mixings, even with fine-tunings. Simply, the radiative effects
that modify the mixings (which are proportional to the tau Yukawa
coupling squared, $y_\tau^2$) are not 
large
enough to do the job for the currently preferred range of masses ($m\leq
0.3$ eV). For the same reason, the mass splittings cannot have a radiative
origin.

\item For the MSSM the amplification is possible but only when (at least
two) neutrinos are quasi-degenerate (in absolute value and sign), and
always involves a fine-tuning between the initial mass splitting (solar or
atmospheric) and its radiative correction: two physically
unrelated quantities that are required to be close to each other. The
magnitude of this fine-tuning is essentially the amplification factor
achieved.
Moreover, a precise value of $\tan\beta$ (very high for the 
atmospheric case) is required.  The
amplification of the atmospheric angle requires an additional and even
more important fine-tuning, since the solar splitting gets a
radiative correction of ``atmospheric'' size which should be compensated 
by an ad-hoc initial condition. Finally, simultaneous amplification of 
solar and atmospheric angles is possible but it is extremely fine-tuned.

All these problems come from the fact that in the SM and the MSSM the mixing 
matrix (when there is some initial quasi-degeneracy) approaches an 
infrared pseudo-fixed-point (IRFP) which implies a physically unacceptable
mixing (solar or atmospheric). Therefore, 
parameters should be delicately chosen for the running
to stop before reaching the IRFP.

\item Things are much better for the scenario of neutrino masses arising 
from the K\"ahler potential. First of all, the infrared fixed points
correspond here to maximal or quasi-maximal mixings, so there is no need 
of fine-tuning in order to amplify angles.
This can work for both the solar and the atmospheric angles.

Moreover, the amplification mechanism can work even if the mass
eigenvalues are not quasi-degenerate. The reason is that the strength
of the RG effect is proportional to $m^A/(m_i-m_j)$, where $m^A$ is the
scale of a coupling that arises from the K\"ahler potential and $m_i$ 
are the
mass eigenvalues [in conventional cases the strength of the effect is 
proportional to
$(m_i+m_j)/(m_i-m_j)$]. So it is enough to have ${m^A\gg m_i-m_j}$
to get important radiative effects, thus reaching the IRFP.
On the other hand, the presence of $m^A$ introduces an additional
uncertainty, which is however removed taking into account that here
the low-energy splitting, $\Delta m_{ij}^2$, is essentially a pure
radiative effect, whose magnitude can be adjusted with the value of $m^A$
without modifying the form of the IRFP.

We find very encouraging that the scenario of neutrino masses from the
K\"ahler potential, which is attractive for different reasons (e.g. it
implies that the scale of lepton number violation is much closer to
the  electroweak scale than in conventional scenarios) has this
remarkable and nice behaviour regarding radiative corrections.

\end{itemize}

To conclude, two more comments are in order. First, radiative effects
play a relevant role in neutrino physics that often cannot be ignored.
E.g. in view of the scarce success of radiative amplification in the MSSM,
one might think that radiative effects are not relevant for model building
in that framework. However, especially for scenarios involving some
quasi-degeneracy, radiative effects can have the (negative) effect of
destabilizing the high-energy pattern of mixing angles and mass 
splittings.
The formulae presented in this paper are useful to analyze
these effects. Second, the radiative corrections studied in this paper
are  model-independent  since they 
concern the running from the $M$-scale (the scale where the new
physics that violates L appears) down to the electroweak scale and this 
running is determined by the effective theory valid in that
range (the SM or the MSSM with a non-renormalizable operator
responsible for neutrino masses). Besides these corrections there
are others arising from physics beyond $M$,\footnote{E.g. 
threshold corrections at $M$ or 
corrections from the
running between a fundamental scale, say $M_P$ or $M_{GUT}$, and $M$. In
the see-saw model this corresponds to the running of the neutrino Yukawa
couplings, $Y_\nu$, and the right-handed neutrino masses between $M_P$ and
$M$. These effects can be quite important (for the see-saw they depends on
the magnitude of $Y_\nu$ and have been analyzed elsewhere 
\cite{CEIN1,Antusch}).} but
they are much more model-dependent. Their role is to set the initial
conditions at $M$ for the model-independent radiative effects analyzed
here.

\section*{A. RGEs for neutrino physical parameters}
\setcounter{equation}{0} 
\renewcommand{\theequation}{A.\arabic{equation}}

The energy-scale evolution of a $3\times 3$ neutrino (Majorana) mass 
matrix ${\cal M}_\nu$ is generically 
described by a RGE of the form ($t=\log Q$): 
\be 
\frac{d {\cal M}_\nu}{dt}=-(u {\cal M}_\nu + P {\cal M}_\nu  + 
{\cal M}_\nu P^T )\ , 
\label{RGMAp} 
\ee 
In (\ref{RGMAp}), $u$ is a number, so $u {\cal M}_\nu$ gives a 
family-universal scaling of ${\cal M}_\nu$ which does
not affect its texture, while $P$ is a matrix in flavour space
thus producing a non family-universal correction
(the most interesting effect).

As explained in \cite{CEIN4} one can extract from (\ref{RGMAp}) 
the RGEs for the physical neutrino parameters: the mass eigenvalues, the 
mixing angles and the CP phases. In this paper we focus for simplicity 
on real cases, with no phases. (General formulas for the case 
with all phases can be found in \cite{CEIN4}.) Using the 
parametrization and conventions of the Introduction, we get
the following RGEs for the mass eigenvalues and the PMNS matrix
\be
\label{RGm}
\frac{d m_i}{dt}= - u m_i - 2 m_i \hat P_{ii}\ ,
\ee
\be
\label{RGV}
\frac{d V}{dt}= V T\ .
\ee
We have defined
\be 
\label{Pdef}
\hat{P}\equiv \frac{1}{2}V^T (P+P^T) V\ ,
\ee
while $T$ is a  $3\times 3$ matrix (anti-hermitian, so that the unitarity 
of $V$ is preserved by the RG running) with 
\bea
\label{Tgendef}
T_{ii}&\equiv&i \hat{Q}_{ii}\ ,\nonumber\vspace{.5cm}\\
T_{ij}&\equiv&
\nabla_{ij} \hat P_{ij}+i\hat{Q}_{ij}\ ,\hspace{1cm} i\neq j\ ,
\eea
where
\be
\hat{Q}\equiv \frac{i}{2}V^T (P-P^T) V\ ,
\ee
and
\be
\label{deltadef}
\nabla_{ij}\equiv\frac{m_i+m_j}{m_i-m_j}.
\ee
Note that the RGE for $V$ does
not depend on the universal factor $u$, as expected.

From eqs.~(\ref{RGV}--\ref{deltadef}), we can derive the 
general RGEs for the mixing angles: 
\bea
\label{RGt1}
\frac{d \theta_1}{dt}&=&
\frac{1}{c_2} \left(
s_3  T_{31} - c_3  T_{32} \right)\ ,\vspace{7mm}\\
\label{RGt2}
\frac{d \theta_2}{dt}&=&- \left(c_3  T_{31} + s_3 T_{32}\right) \ 
,\vspace{7mm}\\
\label{RGt3}
\frac{d \theta_3}{dt}&=&
-\frac{1}{c_2}  T_{21} +
{s_2\over c_2}\left( s_3  T_{31} -
 c_3 T_{32}\right)\ .
\eea
In the next subsections we particularize the generic formulas above to 
scenarios of interest: first to the Standard Model and the MSSM 
and then to models with more sources of lepton number violation, among 
them supersymmetric scenarios with neutrino masses that 
are generated from the K\"ahler potential.

\subsection*{Conventional SM and MSSM}

In the SM or the MSSM the RGE for the neutrino mass matrix (\ref{RGM}) is 
of the form (\ref{RGMAp}). The evolution of neutrino masses is then given 
by (no sum in $i$)
\be
\label{RGmass}
\frac{d m_i}{dt}= - u_M m_i - 2 m_i c_M\hat P_{Eii} \ ,
\ee
where $\hat{P}_E=V^T P_E V$ and $P_{E}\equiv Y_E Y_E^\dagger/(16\pi^2)$ with 
$Y_E$ the matrix of 
leptonic Yukawa couplings, which can be well approximated by $Y_E\simeq 
{\rm diag}(0,0,y_\tau)$ so that 
$16\pi^2\hat P_{Eij}\simeq y_\tau^2 V_{\tau 
i}V_{\tau j}$. The model-dependent quantities $u_M$ and $c_M$
are as follows \cite{RG1}--\cite{RG4}. In the SM
\be
u_M\simeq \frac{1}{16\pi^2}[3g_2^2-2\lambda-6h_t^2]\ , 
\ee
where $g_2,\lambda, h_t$ are the $SU(2)$ gauge
coupling, the quartic Higgs coupling and the top-Yukawa coupling (leptonic 
Yukawa couplings can be safely neglected here), while in the 
MSSM 
\be
u_M=\frac{1}{16\pi^2}\left[
\frac{6}{5}g_1^2+6g_2^2-6\frac{h_t^2}{\sin^2\beta}\right]\ ,
\ee 
where $g_1$ is the $U(1)$ gauge coupling and $\tan\beta$ is the
ratio of the vevs of the two supersymmetric Higgses. Finally,
\be
c_M=\frac{3}{2} 
\hspace{0.5cm} {\rm (SM)}\ ,
\hspace{1.5cm}
c_M=-1
\hspace{0.5cm} {\rm (MSSM)}\ .
\ee 

The RGE for the mixing matrix is of the form (\ref{RGV}) with
\bea 
\label{Tdef}
T_{ii}&=&0\ ,
\nonumber\\
T_{ij}&=&c_M\nabla_{ij}\hat P_{Eij}
\simeq \epsilon_\tau \nabla_{ij} V_{\tau i}V_{\tau j}\ ,
\hspace{1cm} i\neq j \ ,
\eea
with
\be
\label{eps}
\epsilon_\tau\equiv c_M\frac{y_\tau^2}{16\pi^2}\ ,
\ee
where electron and muon Yukawa couplings have been neglected.
The running of the mixing angles, Eqs.~(\ref{RGt1}--\ref{RGt2}), is now 
given by 
($s_i\equiv \sin \theta_i$, etc)
\bea  
\label{teta1}
\frac{d \theta_1}{dt}&=&-\epsilon_\tau c_1 ( -s_3 V_{\tau 1} 
\nabla_{31}+ c_3 V_{\tau 2} \nabla_{32})\ ,  \vspace{7mm}\\ 
\frac{d \theta_2}{dt}&=&-\epsilon_\tau c_1 c_2 ( c_3 V_{\tau 1}  
\nabla_{31}+s_3 V_{\tau 2} \nabla_{32})\ ,  \vspace{7mm}\\
\label{teta3}
\frac{d \theta_3}{dt}&=&-\epsilon_\tau
( c_1 s_2 s_3 V_{\tau 1} \nabla_{31}- c_1 s_2 c_3 V_{\tau 2} 
\nabla_{32} +V_{\tau 1}V_{\tau 2} \nabla_{21})\ .  
\eea  

\subsection*{More general models}

Consider now a $3\times 3$ neutrino mass matrix ${\cal M}_\nu$ that 
evolves with energy following a RGE of the form 
\be
\frac{d {\cal M}_\nu}{dt}=-(u_M
{\cal M}_\nu + c_M  P_E  {\cal M}_\nu + c_M  {\cal M}_\nu P_E^T)
+  P_E M^A_\nu  -   M^A_\nu P_E^T, 
\label{RGEAp}
\ee
where, besides the usual terms there are new contributions coming
from $M^A_\nu$, a $3\times 3$ antisymmetric matrix that arises from 
lepton-violating
physics but is not directly related to neutrino masses. Particular examples 
of RGEs of this form have been considered in the literature 
\cite{Arcadi,PRL,Pantaleone}.

The explicit RGEs for neutrino mass eigenvalues, $m_i$, and mixing matrix,
$V_{\alpha i}$, presented above for generic models with no
$M^A_\nu$-terms [Eqs.~(\ref{RGm}) and (\ref{RGV})] can be immediately 
extended also to Eq.~(\ref{RGEAp}) simply making the substitution 
\be 
P\rightarrow c_M P_E - P_E\ M^A_\nu\ {\cal M}_\nu^{-1}\ ,
\ee 
which transforms (\ref{RGMAp}) into (\ref{RGEAp}). 
More explicitly, we obtain (no sum in $i$)
\be 
\label{RGmassAp} 
\frac{d m_i}{dt}= -u_M m_i  - \frac{y_\alpha^2}{8\pi^2}\left[c_M m_i 
V_{\alpha i}^2 - V_{\alpha i}m^A_{\alpha\beta}V_{\beta i}\right] \ , 
\ee 
and the usual
\be 
\label{RGUAp} 
\frac{d V}{dt}= V T\ , 
\ee 
with $T$ given by 
\bea
\label{TAp} 
T_{ii}&=& 0 \ ,\nonumber\vspace{.5cm}\\ 
16\pi^2T_{ij}&=& c_M y_\alpha^2 V_{\alpha i}V_{\alpha j}\nabla_{ij}-
(y_\alpha^2-y_\beta^2){V_{\alpha i}V_{\beta j}\over 
m_i-m_j}m^A_{\alpha\beta}\ . 
\eea
The explicit RGEs for the mixing angles, Eqs.~(\ref{RGt1}--\ref{RGt3}),
are also valid in this case upon substitution of $T_{ij}$ as above.

If we particularize these results to a case with just 2 flavors 
($\alpha,\beta$), with mass eigenvalues $m_{1,2}$ and
a single mixing angle $\theta$ ($V_{\alpha 2}\equiv \sin\theta$)
we obtain the RGE (no sum in $\alpha,\beta$)
\be
16\pi^2{ d\theta\over dt}={(y_\alpha^2-y_\beta^2)\over m_1-m_2}
\left[{c_M\over 2}(m_1+m_2)\sin 2\theta-m^A_{\alpha\beta}\cos 2\theta
\right]\ .
\ee
{}From this equation we deduce the following. When $M^A_\nu$ is absent, 
there is an infrared pseudo-fixed point at $\theta^*=0,\pi/2$ in the 
evolution of $\theta$. This is the case in most scenarios usually 
considered, for example the SM and the MSSM (with neutrino masses
obtained from the superpotential). On the other extreme case, with 
$c_M$=0 and $M^A_\nu$ present, the pseudo-fixed point is at 
maximal mixing, $\theta^*=\pi/4$. The only instance that we know of 
such case is the MSSM with neutrino masses obtained through the 
K\"ahler potential \cite{PRL} that we have considered in this paper. 
In cases with both $c_M$ and $M^A_\nu$ non-zero, the pseudo-fixed 
point 
is at 
\be
\tan 2\theta^*={2c_M m^A_{\alpha\beta}\over m_1+m_2}\ .
\ee
This latter case can be realized for instance
in non-supersymmetric two Higgs doublet 
models and has been studied previously in Ref.~\cite{Pantaleone}. Note 
however 
that 
the analysis of fixed points in that reference is different from ours: in
\cite{Pantaleone} it is assumed that neutrino masses also reach their fixed 
points, which is not usually the case in most examples of phenomenological 
interest. On the other hand, mixing angles quickly evolve to the fixed 
point if there is quasi-degeneracy of neutrino masses (with same sign 
masses, 
see e.g. the extended discussion in \cite{CEIN4}).

\section*{B. RGEs for couplings in the K\"ahler potential}
\setcounter{equation}{0}
\renewcommand{\theequation}{B.\arabic{equation}}

For a tree-level K\"ahler potential
\be 
K=\sum_a|\phi_a|^2
+{1\over 2 M}\left[\kappa^{ab}_c\phi_a\phi_b\phi^c
+{\mathrm h.c.}\right]
+{1\over 4 M^2}\kappa^{ab}_{cd}\phi_a\phi_b\phi^c\phi^d
+{1\over 6 M^2}\left[\kappa^{abc}_d\phi_a\phi_b\phi_c\phi^d
+{\mathrm h.c.}\right]\ ,
\ee
(with $\phi^a\equiv\phi_a^*$) and a superpotential that includes the 
Yukawa couplings
\be
W=...+{1\over 3!}Y^{abc}\phi_a\phi_b\phi_c\ ,
\ee
one gets, from the condition $dK/dt=0$ and using the one-loop corrected
expression for non-renormalizable K\"ahler potentials computed in
Ref.~\cite{Andrea}, the RGEs
\bea
16\pi^2 {d\kappa^{ab}_c\over dt}&=&
\left[{1\over 2}\kappa^{ax}_cY^{bmn}Y_{xmn}
+2 \kappa^{ax}_yY^{ymb}Y_{xmc}+ (b\leftrightarrow a)\right]
+{1\over 2}\kappa^{ab}_xY^{xmn}Y_{cmn}\nonumber\\
&-&2g_A^2\kappa^{ab}_c
\left[C_2^A(a)+C_2^A(b)-C_2^A(c)\right]\ ,
\eea
\bea
16\pi^2 {d\kappa^{abc}_d\over dt}&=&
\left[{1\over 2}\kappa^{abx}_dY^{cmn}Y_{xmn}
+2 \kappa^{abx}_yY^{cym}Y_{dxm}
- {4\over 3} \kappa^{an}_x\kappa^{bm}_yY^{yxc}Y_{nmd}\right.
\nonumber\\
&-&\left. {4\over 3} \kappa^{an}_x\kappa^{by}_nY^{mxc}Y_{myd}
-{4\over 3} \kappa^{bn}_x\kappa^{ay}_nY^{mxc}Y_{myd}
+ (bca) + (cab)\right]\nonumber\\
&+&{1\over 2}\kappa^{abc}_xY^{xmn}Y_{dmn}
-2g_A^2\kappa^{abc}_d\left[C_2^A(a)+C_2^A(b)+C_2^A(c)-C_2^A(d)\right]\ 
,
\eea
\bea
16\pi^2 {d\kappa^{ab}_{cd}\over dt}&=&
{1\over 2}\left[\kappa^{xb}_{cd}Y^{amn}Y_{xmn}
+(a\leftrightarrow b)\right]
+{1\over 2}\left[\kappa^{ab}_{xd}Y^{xmn}Y_{cmn}
+(c\leftrightarrow d)\right]\nonumber\\
&+&2\left\{\left[\kappa^{xb}_{yd}Y^{amy}Y_{xmc}
-\kappa^{am}_j\kappa^n_{ck}Y^{jkb}Y_{mnd}\right.\right.
-\kappa^{am}_j\kappa^n_{cm}Y^{jkb}Y_{knd}
\nonumber\\
&-&\left.\left.
\kappa^{am}_j\kappa^j_{cn}Y^{nkb}Y_{kmd}
+(c\leftrightarrow d)\right]   
+(a\leftrightarrow b)\right\}\nonumber\\
&+&4g_A^2\left[\kappa^{ab}_{xy}
\left(T_{A\alpha}\right)^x_c\left(T_{A\alpha}\right)^y_d
+\kappa^{xy}_{cd}
\left(T_{A\alpha}\right)^a_x\left(T_{A\alpha}\right)^b_y
\right]\ ,
\eea
where $C_2^A(a)$ is the quadratic Casimir invariant of the chiral field 
$\phi_a$ for the gauge group labelled by $A$ (with gauge coupling $g_A$),
and the $T_{A\alpha}$'s are the group generators (with $A$ labelling the 
group and $\alpha$ the generator). The diagrammatic techniques of 
\cite{Cvitanovic} were useful to compute the non-abelian gauge 
contributions to these equations. 
If the superpotential contains mass terms $\mu^{ab}\phi_a\phi_b$ with 
$\mu^{ab}\sim{\cal O}(\mu)$, the 
previous RGEs would receive additional contributions ({\it e.g.}
terms like $\kappa^{ax}_c\kappa^y_{xz}\mu_{ry}Y^{zrb}$ in the RGE of 
$\kappa_c^{ab}$) proportional to $\mu/M$. We neglect 
them on the assumption that $\mu/M\ll 1$.

\newpage


\end{document}